 \documentclass[12pt]{article}

\usepackage{amsmath}
\usepackage{amsfonts}
\usepackage{amssymb}
\usepackage{tensor}
\input xy
\xyoption{all}
\usepackage{srcltx}

\usepackage{hyperref}

\usepackage{xcolor}

\usepackage[normalem]{ulem}

\usepackage{marginnote}

\input xy
\xyoption{all}

\newtheorem{theo}{\bf \thesection.\arabic{abz}. Theorem }

\newtheorem{lemm}{\bf \thesection.\arabic{abz}. Lemma}

\newtheorem{coro}{\bf \thesection.\arabic{abz}. Corollary}

\newtheorem{defi}{\bf  \thesection.\arabic{abz}. Definition }

\newtheorem{exa}{\bf \thesection.\arabic{abz}. Example}

\newtheorem{rema}{\bf \thesection.\arabic{abz}. Remark}

\newcounter{abz}[section]
\newcounter{equ}[section]
\newcounter{equu}[section]
\newcommand{\abz}{\refstepcounter{abz}}
\newcommand{\equ}{\refstepcounter{equ}}

\def\diag{\mathrm{diag}}

\def\Id{\mathrm{Id}}

\def\F{\mathcal{F}}

\def\Re{\mathop{\mathrm{Re}}}

\def\Span{\mathop{\mathrm{Span}}}

\def\Tr{\mathop{\mathrm{Tr}}}

\def\V{\mathcal{V}}

\def\R{\mathbb{R}}
\def\RR{\mathcal{R}}
\def\L{\mathcal{L}}
\def\P{\mathbb{P}}

\def\al{\alpha}
\def\be{\beta}

\def\ga{\gamma}
\def\de{\delta}

\def\ka{\kappa}
\def\la{\lambda}

\def\om{\omega}

\newcommand{\f}{\mathbf{f}}
\newcommand{\fb}{\bar{f}}

\renewcommand{\t}{\theta}

\newcommand{\xib}{\bar{\xi}}

\newcommand{\iprod}{\mathbin{\lrcorner}}
\def\qed{$\square$}

\providecommand{\bysame}{\leavevmode\hbox to3em{\hrulefill}\thinspace}
\providecommand{\MR}{\relax\ifhmode\unskip\space\fi MR }
\providecommand{\MRhref}[2]{%
  \href{http://www.ams.org/mathscinet-getitem?mr=#1}{#2}
}
\providecommand{\href}[2]{#2}

\title{Dispersionless Hirota system and hidden symmetries of heavenly equation}
\author{
\and
Andriy Panasyuk\\
Faculty of Mathematics and Natural Sciences\\
 Cardinal Wyszy\'{n}ski
University\\
Wóycickiego 1/3, 01-938 Warsaw, Poland\\
\\
Adam Szereszewski\\
Faculty of Physics, University of Warsaw\\
Pasteura 5,  02-093 Warsaw, Poland\\
Adam.Szereszewski@fuw.edu.pl
}
\date{}

\providecommand{\bysame}{\leavevmode\hbox to3em{\hrulefill}\thinspace}
\providecommand{\MR}{\relax\ifhmode\unskip\space\fi MR }
\providecommand{\MRhref}[2]{%
  \href{http://www.ams.org/mathscinet-getitem?mr=#1}{#2}
}
\providecommand{\href}[2]{#2}

\begin{document}
\bibliographystyle{amsalpha}

\maketitle

\begin{abstract}
In 2021 Konopelchenko, Schief and Szereszewski observed that solutions of 4D dispersionless Hirota system also solve the general heavenly equation describing self-dual vacuum Einstein metrics in neutral signature. They also noticed that the symmetry $f\mapsto \Phi(f)$ of the Hirota system essentially changes the properties of the corresponding metric.

In this paper we restate these observations in the context of  I and II Plebański heavenly equation (I,II PHE). Namely, we first find 5D analogues of these  equations. We then consider  a special type of symmetry generalizing the so-called tri-holomorphic symmetry of I or IIPHE. The reduction with respect to this symmetry (which in a sense imitates the reduction of self-dual vacuum Einstein metrics with respect to a tri-holomorphic symmetry ending in special Einstein--Weyl structures) gives an analogue of the dispersionless Hirota system for I and  IIPHE. Such a point of view allows to reinterpret the symmetry $f\mapsto \Phi(f)$ mentioned and obtain explicit formulas for the metric depending on $\Phi$. We present some examples showing how the Weyl spinor changes along with $\Phi$.
\end{abstract}

\tableofcontents
\section*{Introduction}

The classical theory of webs (finite families of foliations on a manifold) was established and intensively developed in the beginning of 20-th century by the school of W. Blaschke in Germany. This study, which was documented in a several dozen of papers and a famous book \cite{blaschkeBol}, has many applications in geometry and differential equations and is continued by many authors all over the world.

The web theory got second wind in the end of that century with the papers of I. Gelfand and I.~ Zakharevich \cite{gz2,gz3,gz4} who used specific 1-parametric families of foliations $\{\F_\la\}$ (now known as Kronecker webs) as a tool for investigation of the local geometry of bihamiltonian structures (pairs of compatible Poisson structures). This interrelation between webs and bihamiltonian structures has been exploited by many authors and led to several local classification results \cite{t1,t2,t7,p1,p2,bouetoudufour}.

Further on, in a seminal paper \cite{z2} I. Zakharevich showed that a subclass of Kronecker webs called Veronese webs  is described in 3D by solutions of a single nonlinear PDE which nowadays is commonly known as dispersionless Hirota equation. In the same paper a twistor theory for this equation was built. The method of Zakharevich can be easily generalized to higher dimensions, where Veronese webs are described by a system of PDEs, which we call the dispesionless Hirota system, see Section \ref{s1}.

In \cite{dunajskiKr,pHirota,KrynskiDef} a series of different integrable nonlinear PDEs related to Veronese webs was established (here by integrability we mean that the PDE possesses a dispersionless Lax pair \cite{BerjFerKrNov}, i.e.
 vector fields $X_1(\la),X_2(\la)$ depending on the second jet of the unknown function $f$ and an auxiliary parameter $\la$ such that the
Frobenius integrability condition $[X_1, X_2] \in \Span(X_1,X_2)$ holds identically modulo the equation and its differential consequences). Also a one-to-one correspondence between solutions of these PDEs and a special 3-dimensional Einstein--Weyl structures was shown.

Another  ingredient of our considerations is the Penrose theory of nonlinear gravitons \cite{penrose}, i.e. self-dual vacuum Einstein  (SDVE) metrics in neutral signature or complex case in dimension 4. It is known \cite{dunajskiBook} that they are in a one-to-one correspondence with Kronecker webs of codimension two with an additional property that the 2-form $\be^\la$ with the quadratic dependence on the parameter $\la$ that annihilates $T\F_\la$ is  closed. In turn, such Kronecker webs are in a one-to-one correspondence with the solutions of the so-called Plebański heavenly equations of I or II type \cite{plebanski}, which are also integrable in the sense mentioned. Many explicit examples of nonlinear gravitons are known in the literature (see \cite{dunajskiBook} and references therein).

Also there are some other integrable PDEs describing gravitons: Husain--Park equation \cite{park}, \cite{husain},
Grant equation \cite{grant},
general heavenly  or Schief equation \cite{schief,doubrovFerapontov,konopSchiefszer}.

Based on the ideas of \cite{pHirota} several new integrable PDEs related to gravitons were constructed and investigated in \cite{pHeavenly}.

One of the novelties in the approach
of \cite{konopSchiefszer} was an observation that among the solutions of the general heavenly equation
there is a subclass of functions satisfying the dispersionless Hirota system of PDEs, i.e. system describing Veronese webs in 4D. In \cite{pHeavenly} this observation was extended to the above mentioned new ``heavenly type'' PDEs and the corresponding ``Hirota type'' systems were calculated explicitly. Geometrically this fact about solutions of the Hirota system and the general heavenly equation means  that the  2-dimensional leaves of the corresponding Kronecker web $\{\F_\la\}$ can be included in the 3-dimensional leaves of the corresponding Veronese web $\{\V_\la\}$.

The relation between  ``Hirota type'' systems of PDEs and the corresponding ``heavenly type'' PDEs in four dimensions is the main topic of the present paper. Its main goal is threefold: (1) to construct the analogs of the ``Hirota type'' systems in 4D corresponding to I and II Plebański heavenly equations, which where not covered by the method of \cite{pHeavenly}; (2) to find some explicit solutions of these systems and use them to construct new explicit SDVE metrics by the ``method of twisting'' that will be described below; (3) to show the uniqueness of the Hirota dispesionless system in a certain sense.

The method  of achieving  the first goal consists in using specific reductions of 5D-members of the heavenly hierarchy of PDEs  to 4D heavenly PDE reminding the well-known reduction of 4D self-dual vacuum Einstein metrics to 3D Einstein--Weyl structures by means of a Killing vector. The main geometric structure which we use for constructing the hierarchy mentioned is a two-form $\be^\la$ which is a polynomial in an auxiliary parameter $\la$ of a certain order, is closed, is a simple tensor, and satisfies some nondegeneracy condition. They were introduced by Gindikin \cite{gindikin82} an we call them the Gindikin structures, see Definition \ref{ginddef}. If a Gindikin structure $\be^\la$ in 5D admits a specific symmetry, i.e. a vector field $K$ such that $\L_K\be^\la=\be^\la$, then there exists a one-form $\al^\la$ with $d\al^\la=\be^\la$ such that after the reduction with respect to $K$ it can serve as the defining form of a Veronese web. The corresponding heavenly system in 5D is reduced to the ``Hirota type'' system in 4D, which in the case of general heavenly hierarchy is the original Hirota system.

The ``method of twisting'' in goal (2) roughly speaking consists  considering instead of $\be^\la=d\al^\la$ the ``twisted'' 2-form $\be_\phi^\la=d(\phi\al^\la)$, where $\phi$ is an arbitrary function of one argument, the last being selected in such a way that $\be_\phi^\la$ is again a Gindikin structure. The corresponding ``twisted'' SVDE metric $g_\phi$ can essentially differ from the initial one $g_1$.

The uniqueness matter we mentioned in goal (3) means the following. We propose a canonical form of both the Gindikin structure $\be^\la$ and the symmetry $K$ in 5D in the ``general heavenly approach'' (see Theorem \ref{th1}) and then show that any $\be^\la$ and any symmetry $K$ can be brought to this form.

The more detailed content of the paper is as follows. In Section \ref{s1} we discuss Veronese webs, dispersionless Hirota system and its symmetries. In particular we show that a general heavenly equation in 4D is an algebraic consequence of the Hirota system. Section \ref{s2} is a short reminder on vacuum self-dual Einstein metrics in neutral signature. In section \ref{gind} we introduce Gindikin structures, their symmetries and discuss how they are related to SDVE metrics in 4D. We also touch the matter of Mason--Newman vector fields which generate the kernel of a Gindikin structure and are divergence free with respect to some volume form. The section is concluded by indicating the Gindikin structures that correspond  to general heavenly, I, and II Plebański heavenly equations.

Section \ref{s4} contains the main results of this paper, Theorems \ref{th1},\ref{th1a}, and \ref{th2}, in which we consider 5D members general heavenly, I, and II Plebański heavenly hierarchies respectively and show how they are related to the corresponding Gindikin structures. Further on, we consider reductions of these 5D heavenly systems by means of a symmetry  of the Gindikin structure and elaborate the reduced systems of PDEs, which serve as analogues of the dispersionless Hirota system in the I and II heavenly cases and coincide with it in the general heavenly case. In each of the cases the original heavenly equation is a member of this Hirota type system.

Section \ref{sTwisted} is devoted to the method of twisting of SDVE metrics that come from the Hirota type systems. We give the formulas for the (inverse) twisted metrics, find some explicit solutions and show how the metric properties (including the fundamental invariants of the Weyl spinor) change after twisting. This discussion is extended to Appendix B, where we consider in detail solutions of the II heavenly Hirota system coming from the so-called $pp$-waves, i.e. obvious solutions of the II Plebański heavenly equation depending on two variables.

In Section \ref{suniq} we solve the question of uniqueness of the Hirota system. Namely, we first show that any 5D Gindikin structure can be put in form (\ref{betaGH}) for appropriate coordinates $x^i$ and a key function $f(x^i)$. Next we show that a symmetry of the Gindikin structure can be always written as $\frac{\partial\ }{\partial x^5}$ and finally we prove that, assuming this symmetry, the function $f$ must obtain a separated form $f=h(x^1,\ldots,x^4)\cdot q(x^5)$. Together with Theorem \ref{th1}, saying in particular that $h$ satisfies the Hirota system in 4D, this guarantees uniqueness of this last.

Appendix \ref{apA} deals with some auxiliary results needed for Section \ref{suniq} that are related to matters of existence in 5D of commuting vector fields spanning $\ker\be^\la$  and a volume form with respect to which they are divergence free. They generalize the above mentioned  Mason--Newman vector fields in 4D.

In Section \ref{sConcl}  some perspectives are outlined.

In the end of this introduction it is worth to mention that there were many attempts to include I and II Plebański heavenly equations into higher dimensional hierarchies \cite{BoyerPlebanski77,BoyerPlebanski85,takasaki,strachan}. In particular, reference \cite{dunMason2000} gives a hierarchy for the II Plebański equation for \emph{even}-dimensional spaces. In our construction of the 5D I or II heavenly system we were led by general principle to describe the Gindikin structure but the methods are different: in the II heavenly case we use twistor function description of the Gindikin structure and in fact our system is a natural generalization of the hierarchy from last reference to dimension 5; in the I heavenly case we use the existence of Mason--Newman vector fields and their generalization to 5D as a guiding principle. In both cases our systems can be supposed to be new.

\section{Veronese webs, dispersionless Hirota system of PDEs, and its $f\to \Phi(f)$ symmetry}
\label{s1}

\abz
\begin{defi}\rm
A \emph{Veronese web} on a manifold $M^{n+1}$ is a family of foliations $\{\F_\la\}_{\la\in\P^1=\R\cup\{\infty\}}$ of codimension 1  which locally is determined  by the annihilating one-form $\al^\la$,  $T\F_\la=\ker\al^\la$, of the form $\al^\la=\al_0+\la\al_1+\cdots+\la^n\al_n$, where $\al_0,\ldots,\al_n$ is a local coframe on $M$ ($\al_n$ annihilates $T\F_\infty$) \cite{gz2}. The terminology is motivated by the fact that the map $\P^1\ni\la\mapsto \Span(\al^\la)\in\P T^*_xM$ is a Veronese curve (rational normal curve).
\end{defi}

In \cite{z2} I. Zakharevich showed that there exists a one-to-one correspondence between Veronese webs in $\R^3$ and equivalence classes of \emph{nondegenerate} solutions to the dispersionless Hirota PDE on a function $f=f(x^1,x^2,x^3)$,
 $$
(\la_2-\la_3)f_1 f_{23}+(\la_3-\la_1)f_2f_{31}+(\la_1-\la_2)f_3f_{12}=0,
 $$
where $\la_i$ are arbitrary pairwise distinct parameters (we write $f_i= \frac{\partial f}{\partial x^i}$ and $f_{ij}= \frac{\partial f}{\partial x^ix^j}$) and by the \emph{nondegeneracy of solution} we mean nonvanishing of $f_i$ for all $i$ (the above mentioned equivalence of solutions will be discussed in the end of this section). The corresponding annihilating 1-form given by
$$
\al^\la=(\la-\la_1)(\la-\la_2)(\la-\la_3)\left(\frac{f_1dx^1}{\la-\la_1} +\frac{f_2dx^2}{\la-\la_2}+
\frac{f_3dx^3}{\la-\la_3}\right)
$$
and the PDE above is equivalent to the Frobenius integrability condition $\al^\la\wedge d\al^\la=0$. This result can be easily generalized to higher dimensions.  Indeed, in $\R^n$ the formula
\begin{equation}\label{falpha}\equ
\al^\la:=\prod_{i=1}^n(\la-\la_i)\sum_{i=1}^n\frac{(\la_{n+1}-\la_i)f_{i}dx^i}{(\la-\la_i)}
\end{equation}
describes a unique (in the projective sense) rational normal curve  of degree $n$ in $\la$ such that for any $i=1,\ldots,n$ we have  $\ker\al^{\la_i}=\ker dx^i$ and $\ker \al^{\la_{n+1}}=\ker df$ ($\la_{n+1}=\infty$ in the formula of Zakharevich). The  Frobenius integrability condition now appears as a system of PDEs equivalent to vanishing of the coefficients of the 3-form $\al^\la\wedge d\al^\la$.

For instance in 4D the corresponding dispersionless Hirota system  is
 \begin{align}\label{hiro}\equ
\nonumber f_1f_{23}(\la_1-\la_5)(\la_2-\la_3)+f_2f_{31}(\la_2-\la_5)(\la_3-\la_1)+f_3f_{12}(\la_3-\la_5)(\la_1-\la_2)&=0\\
\nonumber f_1f_{24}(\la_1-\la_5)(\la_2-\la_4)+f_2f_{41}(\la_2-\la_5)(\la_4-\la_1)+f_4f_{12}(\la_4-\la_5)(\la_1-\la_2)&=0\\
\nonumber f_1f_{34}(\la_1-\la_5)(\la_3-\la_4)+f_3f_{41}(\la_3-\la_5)(\la_4-\la_1)+f_4f_{13}(\la_4-\la_5)(\la_1-\la_3)&=0\\
f_2f_{34}(\la_2-\la_5)(\la_3-\la_4)+f_3f_{42}(\la_3-\la_5)(\la_4-\la_2)+f_4f_{23}(\la_4-\la_5)(\la_2-\la_3)&=0.
 \end{align}
It is easy to see that in fact this system is algebraically dependent. For instance, $eq[4]=\frac1{f_1(\la_1-\la_5)}(f_4(\la_4-\la_5)eq[1]-f_3(\la_3-\la_5)eq[2]+f_2(\la_2-\la_5)eq[3])$, where by $eq[i]$ we denote the corresponding equation. Consequently, the matrix  of system (\ref{hiro}) understood as a system of linear equations on variables $(f_4,f_3,f_2,f_1)$ should be degenerate. This matrix is equal to $M\cdot D$, where $M$ is a skew-symmetric matrix, and $D$ is a diagonal one, $D=\diag(-(\la_5-\la_4),\la_5-\la_3,-(\la_5-\la_2),\la_5-\la_1)$. The Pfaffian of the  skew-symmetric matrix $M$ up to a constant factor is equal to the LHS of the general heavenly equation \cite{schief,doubrovFerapontov,konopSchiefszer}
\begin{equation}\label{heav}\equ
(\la_1-\la_2)(\la_3-\la_4)f_{12}f_{34}-(\la_1-\la_3)(\la_2-\la_4)f_{13}f_{24}
+(\la_1-\la_4)(\la_2-\la_3)f_{14}f_{23}=0.
\end{equation}
This last is therefore algebraically dependent on equations (\ref{hiro})\footnote{Below in Section \ref{gind} we shall also show that equation (\ref{heav}) is also a differential consequence of system (\ref{hiro})}. Moreover, one can show that any one of five  equations (\ref{hiro}--\ref{heav}) is an algebraic consequence of any three of them.

\abz\label{fullhiro}
\begin{defi}\rm
The system of PDEs consisting of equations (\ref{hiro}--\ref{heav}) will be called the \emph{full 4D Hirota system}.
\end{defi}

From the considerations above (since the substitution $f\to \Phi(f)$ only changes the first integral of the foliation $f=const$ not changing the foliation itself) it is easy to deduce that once $f$ is a solution of the full Hirota system, then so is $\Phi(f)$, where $\Phi$ is any smooth function of one variable. One can also check this fact directly.

Moreover, for any fixed $i$ one can substitute $x^i$ by $\phi_i(x^i)$, with $\phi_i$ being a function of a single argument, and this will not change the foliation $x^i=const$. Thus two solutions $f(x^1,\ldots, x^n)$ and $\Phi(f(\phi_1(x^1),\ldots,\phi_n(n^n))$ describe the same Veronese web. This makes evident the equivalence of solutions mentioned in the context of the Zakharevich theorem.

\section{Short reminder on vacuum self-dual Einstein metrics in neutral signature}
\label{s2}

Let $(M^4,g)$ be a pseudo-Riemannian manifold of neutral signature $(++--)$. Then the Hodge star operator $*:\Lambda^2 T^*M\to \Lambda^2 T^*M$ is an involution and there is a decomposition $\Lambda^2 T^*M=\Lambda^2_+ T^*M\oplus \Lambda^2_- T^*M$ to the eigenspaces corresponding to $\pm 1$ eigenvalues. The elements of $\Lambda^2_\pm T^*M$ are called \emph{(anti) self-dual 2-forms}. The Riemann tensor $\RR$ with the symmetries $\RR_{abcd} = \RR_{[ab][cd]}$
can be represented as a linear map
$$
\RR:\Lambda^2 T^*M\to \Lambda^2 T^*M
$$
and in matrix form corresponding to the decomposition above as
$$
\RR=\left[\begin{array}{cc}
C_+-\frac{R}{12} & \psi\\
\psi & C_--\frac{R}{12}
\end{array}\right],
$$
where $\psi$ is the traceless Ricci tensor, $C_{\pm}$ is the (anti) self-dual Weyl tensor, and $R$ is the Ricci scalar.

\abz
\begin{defi}\rm
{\em A self-dual vacuum Einstein (SDVE) metric} is a metric $g$ such that $\psi=0$, $C_-=0$ and $R=0$, i.e.
$$
\RR=\left[\begin{array}{cc}
C_+ & 0\\
0 & 0
\end{array}\right].
$$
\end{defi}
The terminology is due to the fact that  such a metric satisfies the vacuum Einstein equation.

\section{Gindikin structures, Mason--Newman vector fields, SDVE metrics, and their symmetries}
\label{gind}
\abz\label{ginddef}
\begin{defi}\rm (\cite{gindikin82}) An \emph{$E$-structure} of degree $k$ is a polynomially depending on an auxiliary  parameter $\la$ two-form $\be^\la=\be_0+\la\be_1+\cdots+\la^k\be_k$ on a manifold $M$ such that
\begin{itemize}\item $d\be^\la\equiv0$;
\item $\be^\la\wedge\be^\la\equiv0$.
\end{itemize}
Say that an {$E$-structure} $\be^\la$ is nondegenerate (or simply a \emph{Gindikin structure}) if
\begin{itemize}\item $\deg\be^\la=\dim M-2$;
 \item $\be^{\la_1}\wedge\cdots\wedge\be^{\la_n}\not=0$ for any pairwise different $\la_1,\ldots,\la_n$,  $n=[\dim M/2]$.
 \end{itemize}
\end{defi}

Using the Grothendick theorem on decomposition of vector bundles on a rational normal curve one can prove that, given a  Gindikin  structure $\be^\la$ on $\R^N$, there exist  1-forms $\ga_0,\ldots, \ga_{n-1}, \de_0,\ldots,\de_{k-1}$ forming a local coframe, where $k=n$ for $N=2n$ or $k=n-1$ for $N=2n-1$,   such that $\be^\la=(\ga_0+\cdots+ \la^{n-1}\ga_{n-1})\wedge(\de_0+\cdots+ \la^{k-1}\de_{k-1})$ (the condition $d\be^\la=0$ gives additional restrictions on these 1-forms). In particular, we have
\begin{itemize}\item  $\be^\la=(\ga_0+\la\ga_1)\wedge(\de_0+\la\de_1)$ in 4D (an elementary proof of this fact was given in \cite{gindikin82})
\item $\be^\la=(\ga_0+\la\ga_1+\la^2\ga_2)\wedge(\de_0+\la\de_1)$ in 5D
\end{itemize}
for some linearly independent 1-forms $\ga_0,\ldots,\de_1$.

\abz\label{gii}
\begin{theo}(\cite{gindikin82}) There is a 1-1-correspondence between SDVE metrics and nondegenerate $E$ structures in 4D.
\end{theo}
One side of this correspondence is  as follows: given
$$
\be^\la=(\ga_0+\la\ga_1)\wedge(\de_0+\la\de_1),
$$
the corresponding metric is written as
$$
g=\det\left[\begin{array}{cc}
\ga_0 & \ga_1\\
\de_0 & \de_1
\end{array}\right]=
\ga_0\odot\de_1-\ga_1\odot\de_0.
$$

Algebraically, there are two $SL(2)$ actions on $\ga^{1:\la}=1 \ga_0+\la\ga_1$, $\de^{1:\la}=1 \de_0+\la\de_1$:
\begin{itemize}\item $\ga^{1:\la}\mapsto \ga^{a+b\la:c+d\la}$, $\de^{1:\la}\mapsto \de^{a+b\la:c+d\la}$
\item $\ga^{1:\la}\mapsto a \ga^{1:\la}+b \de^{1:\la}$, $\de^{1:\la}\mapsto c \ga^{1:\la}+d \de^{1:\la}$, \end{itemize}
 where $\left[\begin{array}{cc}
a & b\\
c & d
\end{array}\right]\in SL(2)$. Both of them preserve $g$.

Geometrically, there is a unique  up to a factor contravariant metric $\tilde{g}$ such that the plane $\Span(\ga^{1:\la},\de^{1:\la})$ lies in the null cone of $\tilde{g}$ for any $\la$. The inverse metric $\tilde{g}^{-1}$ up to a factor coincides with $g$. The condition of closedness of $\be^\la$ is equivalent to the ``half-flatness'' of the metric.

Vice versa, any metric can be put locally in the form
$$
g=\det\left[\begin{array}{cc}
\ga_0 & \ga_1\\
\de_0 & \de_1
\end{array}\right]
$$
for some local coframe $\ga_0,\ldots,\de_1$ and a gauge freedom remains, $g\mapsto XgY$, where $X,Y\in SL(2)$, from which $Y$ preserves
$$
\be^\la:=(\ga_0+\la\ga_1)\wedge(\de_0+\la\de_1)=:\be_0+\la \be_1+\la^2\be_2
$$
and $X$  projectively changes $\la$, which gives a natural $SL(2)$-action on the space $\Span(\be_0,\be_1,\be_2)$ of self-dual 2-forms.

Given a conformal Killing vector $K$, $\L_Kg=cg$, of a SDVE metric $g$, $c$ is necessary a constant
and $K$ preserves the space of self-dual forms. In particular, if any of these forms are preserved individually, i.e.
\begin{equation}\label{trihol}\equ
\L_K\be^\la=c\be^\la,\ \mbox{or,\ equivalently, } \ \L_K\be_i=c\be_i,
\end{equation}
then $K$ is the so-called \emph{triholomorphic symmetry} of $g$ \cite{dunajskiMason}.

In the following definition we generalize the notion of triholomorphic symmetry of a SDVE metric, mimicking formula (\ref{trihol}).

\abz\label{symm}
\begin{defi}\rm Given a Gindikin structure $\be^\la=\sum\la^i\be_i$, we say that a vector field $K$ is its symmetry if there exists a constant $c$ such that
\begin{equation}\label{eqsym}\equ
\L_K\be^\la\equiv c\be^\la,
\end{equation}
or, equivalently,
$$
\L_K\be_i=c\be_i
$$
for any $i$, where $\L_K$ stands for the Lie derivative along $K$.
\end{defi}

It is worth mentioning yet another point of view at SDVE metrics. In 1989 Mason and Newman \cite{masonNewman} proved that there is a 1-1-correspondence between SDVE metrics  and pairs of commuting parameter depending vector fields $X_1(\lambda),X_2(\lambda)$ which are divergence free with respect to some volume form. If \begin{equation}\label{intro1}\equ
 X_1(\lambda)=V_1+\la V_2, \quad X_2(\lambda)=V_3+\la V_4, \quad [X_1(\la),X_2(\la)]\equiv 0, \quad \mathcal{L}_{V_i}\om=0
 \end{equation}
 for some volume form $\om$, the metric
 \begin{equation}\label{intro2}\equ
 g=\kappa(V^1\odot V^4-V^2\odot V^3),
 \end{equation}
 where $\kappa:=\om(V_1,V_2,V_3,V_4)$ and $(V^i)$ stands for the coframe dual to the frame $(V_i)$, is SDVE. The corresponding Gindikin structure now appears as $\be^\la:=X_1(\la)\iprod (X_2(\la)\iprod \om)$ and $X_1(\lambda),X_2(\lambda)$ span the 2-dimensional kernel of $\be^\la$ (cf. \cite{pHeavenly} and Appendix \ref{apA}).

We finish this section by indicating explicitly the Gindikin structures and the corresponding SDVE metrics in the formalisms of general heavenly, I, and II Plebański heavenly equations.

\medskip

\noindent\underline{General heavenly equation.} Consider the 2-form
\begin{equation}\label{betaGH}\equ
\be^\la:=(\la-\la_1)\cdots(\la-\la_{n})\sum_{1\le i<j\le n}\frac{(\la_i-\la_j)f_{ij}dx^i\wedge dx^j}{(\la-\la_i)(\la-\la_j)}
\end{equation}
on $\R^n$.
It is easy to check that
\begin{equation}\label{betaGHa}\equ
\be^\la=\frac1{\la-\la_{n+1}}d\al^\la
\end{equation}
with $\al^\la$ given by (\ref{falpha}).  For $n=4$ the condition $\be^\la\wedge\be^\la\equiv 0$ is equivalent to the general heavenly PDE (\ref{heav}).

Now we can also establish the fact that equation (\ref{heav}) is a differential consequence of system (\ref{hiro}), which follows from the obvious implication $\al^\la\wedge d\al^\la\equiv 0\Longrightarrow d\al^\la\wedge d\al^\la\equiv 0$.

We see that the Gindikin structure corresponding to the general heavenly equation, as well as the equation itself, is intimately related to ``Veronese distribution'' annihilated by the 1-form $\al^\la$.

The matrix of the corresponding metric $g$ in $(dx^1,\ldots,dx^4)$-basis is given by
$$
 [g]:=\frac1{J}\left[\begin{array}{cccc}
2f_{12}f_{13}f_{14} & f_{12}(f_{14}f_{23}+f_{13}f_{24}) & f_{13}(f_{14}f_{23}+f_{12}f_{34}) & f_{14}(f_{13}f_{24}+f_{12}f_{34})\\
* & 2 f_{12}f_{23}f_{24} & f_{23}(f_{13}f_{24}+f_{12}f_{34}) & f_{24}(f_{14}f_{23}+f_{12}f_{34}) \\
* & * & 2f_{13}f_{23}f_{34} & f_{34}(f_{14}f_{23}+f_{13}f_{24})\\
* & * & * & 2f_{14}f_{24}f_{34}
\end{array}\right],
 $$
where $J=f_{14}f_{23}-f_{13}f_{24}$ and the asterisks are used to shorten the form of a symmetric matrix.
\bigskip

\noindent\underline{I Plebański equation} We pass to the coordinates $(r,s,z,w)$. Here the Gindikin structure is given by the formulas
\begin{align}\label{gindIH}\equ
\be^\la&=\la^2dr\wedge ds+\la(\theta_{rz}dr\wedge dz+\theta_{rw}dr\wedge dw+\theta_{sz}ds\wedge dz+\theta_{sw}ds\wedge dw)-dz\wedge dw=\ga^\la\wedge \de^\la,\\ \nonumber
\ga^\la&=dw+\la(\theta_{rz}dr+\theta_{sz}ds),\de^\la=dz-\la(\theta_{rw}dr+\theta_{sw}ds),
\end{align}
where the key function $\theta$ satisfies the  I Plebański heavenly PDE $\theta_{sz}\theta_{rw}-\theta_{rz}\theta_{sw}=1$ that guarantees the equality $\be^\la=\ga^\la\wedge \de^\la$.

This Gindikin structure can be characterized by the  Mason--Newman vector fields
$$
X_1(\lambda)= \la\partial_w+(\theta_{sw}\partial_r -\theta_{rw}\partial_s),X_2(\lambda)= \la\partial_z+(\theta_{sz}\partial_r-\theta_{rz}\partial_s),
$$
which are divergence free with respect to the standard volume form $\om=dr\wedge ds\wedge dz\wedge dw$ (cf. \cite{dunMason2000}). The closedness of $\be^\la=X_1(\la)\iprod (X_2(\la)\iprod \om)$ is a byproduct. The corresponding SDVE metric is given by
\begin{align}\label{metrIH}\equ
g&=-(\theta_{rz}dr dz+\theta_{rw}dr dw+\theta_{sz}ds dz+\theta_{sw}ds dw)\nonumber\\
&=-(\theta_{rw}dr+\theta_{sw}ds)(dw)-(dz) (\theta_{rz}dr+\theta_{sz}ds).
\end{align}

\medskip

\noindent\underline{II Plebański equation}
We shall use the coordinates $(x,y,z,w)$. In this case the Gindikin structure is related to the so-called twistor functions, i.e. analytic in $\la$ functions that can be pulled back from the coordinate space to the correspondence space and then descend to functions on the twistor space. More precisely, these two twistor functions depend on $x,y,z,w$ and the jet of the key-function and correspond to the Darboux coordinates of the canonical symplectic form on the twistor space \cite[Ch. 10.5]{dunajskiBook}. The Gindikin structure $\be^\la$ appears as the truncation of the following expression to order 2:
\begin{align}\label{GSIIH}\equ
 d(w+\la y-\la^2\theta_x+\la^3\theta_z+\cdots)\wedge d(z-\la x-\la^2\theta_y-\la^3\theta_w+\cdots)\nonumber\\
=dw\wedge dz+\la(dx \wedge dw + dy \wedge dz) + \la^2( dx\wedge dy - dw \wedge d(\theta_y) + dz \wedge d(  \theta_x ))+o(\la^2)=:\be^\la+o(\la^2).
\end{align}
Here the h.o. terms $o(\la^2)$ vanish if the II Plebański heavenly equation
\begin{equation}\label{IIPL}\equ
\theta_{xx}\theta_{yy}-\theta_{xy}^2+\theta_{xw}+\theta_{yz}=0
\end{equation}
is satisfied by the key function $\theta(x,y,z,w)$. This equation is also equivalent to the condition $\be^\la\wedge\be^\la\equiv 0$.

The corresponding metric is given by
\begin{equation}\label{metrIIH}\equ
g=dwdx + dzdy -  \theta_{xx} dz^2 - \theta_{yy} dw^2 + 2  \theta_{xy} dwdz.
\end{equation}
In this formalism the Weyl spinor of $g$, which reflects the only nonvanishing part of the curvature, has especially simple form:
 $C_{ABCD} =\frac{\partial^4\theta}{\partial x^A \partial x^B \partial x^C \partial x^D}$, where $x^A=(y,-x)$.

\abz \label{refnond}
\begin{rema}\rm The condition of the nondegeneracy $\be^\la\wedge\be^\mu\not= 0$ of the Gindikin structures $\be^\la$ for the above mentioned cases of general heavenly, I, and II Plebański heavenly equations is satisfied generically. Indeed, the condition $\be^\la\wedge\be^\mu= 0$ for $\la\not=\mu$ is equivalent to additional PDE on the key function, which need not be satisfied for generic solutions of the corresponding equation. The same considerations apply to 5D Gindikin structures, which will appear below.
\end{rema}

In the next  section we shall  construct the generalizations of the corresponding Gindikin structures to 5D and consider their particular symmetries.

\section{5D analogues of general, I, and II Plebański heavenly equations and their reductions}
\label{s4}

For each of    the three equations  mentioned in the title of this section our motivation for generalization to 5D is different. We start from the general heavenly case.

The original considerations of W. Schief, which have led to the study of relations between  Hirota system (\ref{hiro}) and  general heavenly equation (\ref{heav}), were as follows. Take the ``5D \emph{general heavenly system}''\footnote{It can be shown that only 3 of 5 equations of this system are algebraically independent.}
 \begin{align}\label{schief}\equ
\nonumber(\la_1-\la_2)(\la_3-\la_4)f_{12}f_{34}-(\la_1-\la_3)(\la_2-\la_4)f_{13}f_{24}
+(\la_1-\la_4)(\la_2-\la_3)f_{14}f_{23}=0\\
\nonumber(\la_1-\la_2)(\la_3-\la_5)f_{12}f_{35}-(\la_1-\la_3)(\la_2-\la_5)f_{13}f_{25}
+(\la_1-\la_5)(\la_2-\la_3)f_{15}f_{23}=0\\
\nonumber(\la_1-\la_2)(\la_4-\la_5)f_{12}f_{45}-(\la_1-\la_4)(\la_2-\la_5)f_{14}f_{25}
+(\la_1-\la_5)(\la_2-\la_4)f_{15}f_{24}=0\\
\nonumber(\la_1-\la_3)(\la_4-\la_5)f_{13}f_{45}-(\la_1-\la_4)(\la_3-\la_5)f_{14}f_{35}
+(\la_1-\la_5)(\la_3-\la_4)f_{15}f_{34}=0\\
(\la_2-\la_3)(\la_4-\la_5)f_{23}f_{45}-(\la_2-\la_4)(\la_3-\la_5)f_{24}f_{35}
+(\la_2-\la_5)(\la_3-\la_4)f_{25}f_{34}=0.
\end{align}
It appears as a natural 5D generalization of general heavenly equation (\ref{heav}) based on the symmetric shape of this last.
Substituting $f(x^1,x^2,x^3,x^4,x^5)=h(x^1,x^2,x^3,x^4)\cdot q(x^5)$ one gets, in a vicinity of points with  $q(x_5),q'(x_5)\not=0$,  the full 4D Hirota system on the function $h$ (see Definition \ref{fullhiro}). We shall put this observation in the geometric context and generalize it to the framework of I and II Plebański heavenly equations.

\abz\label{th1}
\begin{theo}
\begin{enumerate}\item  ``5D general heavenly system'' (\ref{schief}) is equivalent to condition $\be^\la\wedge\be^\la\equiv 0$ for the 2-form $\be^\la$ given by formula (\ref{betaGH}) with $n=5$. Consequently, due to (\ref{betaGHa}) this 2-form is closed and is a Gindikin structure if the function $f$ satisfies (\ref{schief}), see also Remark \ref{refnond}.
\item Assume that the  function $f(x^1,x^2,x^3,x^4,x^5)=h(x^1,x^2,x^3,x^4)\cdot q(x^5)$ satisfies system (\ref{schief}), where $h$ and $q$ are smooth functions of the corresponding arguments and $q'(x^5)\not=0$. Then
\begin{enumerate}
\item the  vector field $K=\frac{q(x^5)}{q'(x^5)}\frac{\partial}{\partial x^5}$ is a symmetry of the corresponding Gindikin structure  with $c=1$, see Definition \ref{symm};
\item $\be^\la=d\al^\la$, where $\al^\la:=K\iprod \be^\la$;
\item
on each hypersurface $S$ given by $x^5=const$ the restriction $\be^\la|_S$ of $\be^\la$ is a 4D Gindikin structure; moreover, $\be^\la|_S=d(\al^\la|_S)$ and, under an additional assumption of nondegeneracy of $f|_S$ (which means $\frac{\partial h(x)}{\partial x^i}\not=0$, $i=1,\ldots,4$, and $q(x^5)\not=0$ on $S$), the 1-form $\al^\la|_S$ defines a Veronese web on $S$;
\item the function $h(x^1,x^2,x^3,x^4)$ satisfies a system of PDEs depending on $\la_5$ coinciding with  the full 4D Hirota system (see Definition \ref{fullhiro}) and $\be^\la|_S$ is the Gindikin structure of the corresponding SDVE metric.
\end{enumerate}
\end{enumerate}
\end{theo}

\noindent{\sc Proof} Item 1 and  Item 2(a) are proved by the direct inspection.  Item 2(b) follows from the Cartan formula for the Lie derivative. The restriction of $\be^\la$ to $S$  coincides with the 2-form (\ref{betaGH}) with $n=4$ and $f=q(const)\cdot h$. Similarly, the restriction of $\al^\la$ to $S$ up to a constant factor coincides with the 1-form (\ref{falpha}) with $n=4$ and $f=h$, which proves Item 2(c). Finally, Item 2(d) follows from  the preceding item and considerations of Sections \ref{s1} and \ref{gind}. \qed

\medskip

Our next aim is to find a 5D analogue of  4D Gindikin structure $\be^\la$ given by formula (\ref{gindIH}) corresponding to I Plebański heavenly equation and then to generalize the reduction procedure described in Item 2 of Theorem \ref{th1}. In the first task we shall be guided by the fact that $\be^\la=X_1(\la)\iprod (X_2(\la)\iprod \om)$ for some commuting vector fields $X_1(\la),X_2(\la)$ linearly depending on $\la$ which are divergence free with respect to the standard coordinate volume form $\om$.

Consider the  2-form $\be^\la$ and 1-forms $\ga^\la,\de^\la$  given in coordinates $(r,s,z,w,u)$ by
\begin{align}\label{gindIH5D}\equ
\be^\la&=\la^3 dr\wedge ds+\la^2(\Theta_{rz}dr\wedge dz+\Theta_{rw}dr\wedge dw+\Theta_{ru}dr\wedge du+\Theta_{sz}ds\wedge dz+\Theta_{sw}ds\wedge dw\nonumber\\
&+\Theta_{su}ds\wedge du)
+\la(\Theta_{rw}dr\wedge du+\Theta_{sw}ds\wedge du-dz\wedge dw+\Theta_{zw}dz\wedge du+\Theta_{ww}dw\wedge du)
-dz \wedge du,\\
\ga^\la&=\la^2(\Theta_{rz}dr+\Theta_{sz}ds)+\la(dw-\Theta_{zw}du)+du\nonumber\\
\nonumber
\de^\la&=\la(-\Theta_{rw}dr-\Theta_{sw}ds)+dz+\Theta_{ww}du,
\end{align}
a triple of $\la$-depending vector fields
\begin{align}\label{MNIH5D}\equ
X_1(\la) &=\Theta_{su}\partial_r- \Theta_{ru}\partial_s-\partial_w+\la\partial_u;\nonumber\\
X_2(\la) &= \Theta_{s w}\partial_r- \Theta_{r w}\partial_s+\la\partial_w;\nonumber\\
X_3(\la) &= \Theta_{s z}\partial_r- \Theta_{r z}\partial_s+\la\partial_z,
\end{align}
and the following system of PDEs on the function $\Theta(r,s,z,w,u)$, ``5D I heavenly system'':
\begin{align}\label{IH5D}\equ
\Theta_{sz}\Theta_{rw}-\Theta_{rz}\Theta_{sw}=1;\nonumber\\
\Theta_{rw}\Theta_{zw}-\Theta_{rz}\Theta_{ww}+\Theta_{ru}=0;\nonumber\\
\Theta_{sw}\Theta_{zw}-\Theta_{sz}\Theta_{ww}+\Theta_{su}=0.
\end{align}

\abz\label{th1a}
\begin{theo}
\begin{enumerate}\item The vector fields $X_i(\la)$, $i=1,2,3$ are divergence free with respect to the standard coordinate volume forme $\om=dr\wedge ds\wedge dz\wedge dw \wedge du$.
\item A function $\Theta(r,s,z,w,u)$ satisfies system (\ref{IH5D}) if and only if $\be^\la=\ga^\la\wedge\de^\la$.
\item A function $\Theta(r,s,z,w,u)$ satisfies system (\ref{IH5D}) if and only if $\be^\la=X_1(\la)\iprod (X_2(\la)\iprod (X_3(\la)\iprod \om))$.
\item If the function $\Theta(r,s,z,w,u)$ is of class $C^3(\R^5)$ and  satisfies system (\ref{IH5D}), then the vector fields $X_i(\la)$, $i=1,2,3$ pairwise commute and the 2-form $\be^\la$ is closed (and consequently is a Gindikin structure).
\item Assume that system (\ref{IH5D}) is satisfied by the key-function of the form $\Theta(r,s,z,w,u)=u\,\theta(r,\frac{s}{u},z,\frac{w}{u})$ with smooth $\theta$.  Then
\begin{enumerate}\item the vector field $K=s{\partial_s}+w{\partial_w}+u{\partial_u}$ satisfies (\ref{eqsym})  with $c=1$.
\item $\be^\la=d\al^\la$, where $\al^\la:=K\iprod \be^\la$ (since $\L_K\be^\la=d(K\iprod \be^\la)$);
\item
on each hypersurface $S$ given by $u=const$ the restriction  $\be^\la|_S$ of $\be^\la$ up to the factor $\la$ is a 4D Gindikin structure; moreover,  $\be^\la|_S=d(\al^\la|_S)$ and, under an additional assumption of nondegeneracy\footnote{\label{ftn1} In this case the assumption of nondegeneracy of $\Theta|_S$ cannot be so easily formulated as in the case of general heavenly equation, cf. Item 2(c) of Theorem \ref{th1}, but it is equivalent to the condition that the coefficients of $\al^\la|_S$ in $\la$ form a coframe on $S$.}  of $\Theta|_S$, $\al^\la|_S$ is a 1-form defining a Veronese web on $S$;
\item the function $\theta(r,s/u,z,w/u)$ restricted to $S$ satisfies a system of PDEs depending on the parameter $u$ which for $u=1$ looks as follows:
 \begin{align}\label{IHir4D}\equ
\theta_{sz}\theta_{rw}-\theta_{rz}\theta_{sw}=1;\nonumber\\
\theta_{zw}\theta_{rw}-\theta_{rz}\theta_{ww}-s \theta_{rs}-w\theta_{rw}+\theta_{r}=0;\nonumber\\
\theta_{sw}\theta_{zw}-\theta_{sz}\theta_{ww}-s\theta_{ss}-w\theta_{sw}=0.
\end{align}
Moreover, for $u=1$ the 4D Gindikin structure $\be^\la|_S$  coincides up to the factor $\la$  with that given by (\ref{gindIH}) being the Gindikin structure of the corresponding SDVE metric (\ref{metrIH}). Thus, by analogy with the general heavenly case, system (\ref{IHir4D}) can be interpreted as the full Hirota system in the I heavenly framework.
\end{enumerate}
\end{enumerate}
\end{theo}

\noindent{\sc Proof} Items 1 and 2 are proved by direct inspection. In the proof of Item 3 one uses the following two PDEs
\begin{align}\label{IHadd}\equ
\Theta_{sz}\Theta_{ru}-\Theta_{rz}\Theta_{su}+\Theta_{zw}=0;\nonumber\\
\Theta_{sw}\Theta_{ru}-\Theta_{su}\Theta_{rw}+\Theta_{ww}=0,
\end{align}
which are the algebraic consequences of (\ref{IH5D}). Namely first of them is equal to $eq[3] \Theta_{rz}-eq[2] \Theta_{sz}$ modulo $eq[1]$ and the second one to $eq[3] \Theta_{rw}-eq[2] \Theta_{sw}$ modulo $eq[1]$, where $eq[i]$ stands for the $i$-th equation of system (\ref{IH5D}).
The proof of vanishing of commutators in Item 4 uses first partial derivatives of $eq[1]$ and of equations (\ref{IHadd}). The closedness of the 2-form $\be^\la$ follows from \cite[Th. 2.2] {pHeavenly}.

Item 5(b) follows from the Cartan formula for the Lie derivative and from Item 5(a). The rest of the proof is performed by direct check.
\qed

\abz
\begin{rema}\rm
 An explicit formula for the 1-form $\al^\la|_S$, where $S=\{u=1\}$,  will be used below:
 \begin{align}\label{alIHS}\equ
dz+\la(-\theta_{rw}dr-\theta_{sw}ds+(-\theta_{zw}+w)dz-\theta_{ww}dw)\nonumber\\
+\la^2((s\theta_{rs}-\theta_{r})dr+
s\theta_{ss}ds+s\theta_{sz}dz+s\theta_{sw}dw)
-\la^3sdr.
\end{align}
One can check directly that equations (\ref{IHir4D}) are equivalent to the Frobenius integrability condition $(\al^\la|_S)\wedge d(\al^\la|_S)=0$.
\end{rema}

\abz
\begin{rema}\rm
It is worth to mention that an attempt to write down an analogue of the Hirota system in frames of the I Plebański heavenly equation was made in \cite[Sec. 11]{konopSchiefszer} based on a (quite nontrivial) passage from the general heavenly equation to the I Plebański one. As a result one gets a system of PDEs \emph{not} coinciding with (\ref{IHir4D}). Note however,  that in view of matter of uniqueness of the Hirota system discussed in Section \ref{suniq} these two systems should be equivalent. Unfortunately, we are not able to find an explicit transformation between them.
\end{rema}

\medskip

To formulate an analogue of Theorems \ref{th1}, \ref{th1a} for the II Plebański equation consider the following $\la$-depending 2-form in 5 variables $(x,y,z,w,u)$ with the key-function $\Theta$:
\begin{align}\label{GSIIP}\equ
\be^\la=du\wedge dz+\la(dx \wedge du + dw \wedge dz) + \la^2(dx\wedge dw+ dy\wedge dz - du \wedge d(\Theta_y))\nonumber \\
+\la^3( dx\wedge dy + dz \wedge d(\Theta_x) - dw \wedge d(  \Theta_y )- du \wedge d(\Theta_w)).
\end{align}
Note that up to the terms of  order $\ge 3$ in $\la$ this form coincides with the form (cf. formula (\ref{GSIIH}))
\begin{equation}\label{GSsympl}\equ
 d(u+\la w+\la^2 y-\la^3\Theta_x+\cdots)\wedge d(z-\la x-\la^2\Theta_y-\la^3\Theta_w+\cdots),
 \end{equation}
 which implies that $\be^\la$ is closed.

\abz\label{th2}
\begin{theo} Let $\be^\la$ be defined by (\ref{GSIIP}).
\begin{enumerate}
\item
Condition $\be^\la\wedge\be^\la\equiv0$ is equivalent to the following system of PDEs (``5D II heavenly system''):
\begin{align}\label{IIH5D}\equ
\Theta_{xx}\Theta_{yy}-\Theta_{xy}^2+\Theta_{xw}+\Theta_{yz}=0 \nonumber \\
-\Theta_{xw} \Theta_{xy}+\Theta_{xx} \Theta_{yw}+\Theta_{xu}+ \Theta_{zw}=0 \nonumber \\
\Theta_{xw} \Theta_{yy}-\Theta_{xy} \Theta_{yw}+\Theta_{ww}- \Theta_{yu}=0.
\end{align}
Consequently, the 2-form $\be^\la$ is a Gindikin structure if the function $\Theta$ satisfies (\ref{IIH5D}), see also Remark \ref{refnond}.
\item Assume that system (\ref{IIH5D}) is satisfied by the key-function of the form $\Theta(x,y,z,w,u)=u\,\theta(x,\frac{y}{u},z,\frac{w}{u})$ with smooth $\theta$.  Then
\begin{enumerate}\item the vector field $K=y\frac{\partial}{\partial y}+w\frac{\partial}{\partial w}+u\frac{\partial}{\partial u}$ satisfies (\ref{eqsym})  with $c=1$.
\item $\be^\la=d\al^\la$, where $\al^\la:=K\iprod \be^\la$ (since $\L_K\be^\la=d(K\iprod \be^\la)$);
\item
on each hypersurface $S$ given by $u=const$ the restriction  $\be^\la|_S$ of $\be^\la$ up to the factor $\la$ is a 4D Gindikin structure; moreover,  $\be^\la|_S=d(\al^\la|_S)$ and, under an additional assumption of nondegeneracy\footnote{Analogously to the case of I heavenly equation it is equivalent to the condition that the coefficients of $\al^\la|_S$ in $\la$ form a coframe on $S$, cf. footnote \ref{ftn1} on page \pageref{ftn1}.}  of $\Theta|_S$, $\al^\la|_S$ is a 1-form defining a Veronese web on $S$;
\item the function $\theta(x,y/u,z,w/u)$ restricted to $S$ satisfies a system of PDEs depending on the parameter $u$ which for $u=1$ looks as follows:
 \begin{align}\label{IIH4D}\equ
\theta_{xx}\theta_{yy}-\theta_{xy}^2+\theta_{xw}+\theta_{yz}=0\nonumber \\
-\theta_{xw} \theta_{xy}+\theta_{xx} \theta_{yw}-y\theta_{xy}-w\theta_{wx}+\theta_{wz}+\theta_x=0\nonumber\\
\theta_{xw} \theta_{yy}-\theta_{xy} \theta_{yw}+y\theta_{yy}+w\theta_{wy}+\theta_{ww}=0.
\end{align}
Moreover, for $u=1$ the 4D Gindikin structure $\be^\la|_S$  coincides up to the factor $\la$ with that given by (\ref{GSIIH}) being the Gindikin structure of the corresponding SDVE metric (\ref{metrIIH}). Thus, by analogy with the general heavenly case, system (\ref{IIH4D}) can be interpreted as the full Hirota system in the II heavenly framework.
\end{enumerate}
\end{enumerate}

\end{theo}

\noindent{\sc Proof} Item 2(b) follows from the Cartan formula for the Lie derivative and from Item 2(a). The rest of the proof similarly to the preceding theorem  is performed by direct check. \qed

\abz
\begin{rema}\rm
Gindikin structure (\ref{GSIIP}) is a natural extension  to 5D of the Gindikin structure related to the so-called hyper-K\"{a}hler hierarchy of Dunajski--Mason, which is defined in any \textit{even} dimension \cite{dunMason2000,dunajskiMason}. The symmetry $K$ above is a generalization of the triholomorphic symmetry of a SDVE metric leading to Lorentzian hyper-CR Einstein--Weyl structures \cite{dunajski}.
\end{rema}

\abz
\begin{rema}\rm
 We would like to give an explicit formula for the 1-form $\al^\la|_S$ which will be used below. Notice that due to the obvious relations
$$
\L_Kz=\L_Kx=\L_K\Theta_y=\L_K\Theta_w=0,\L_Ku=u,\L_Kw=w,\L_Ky=y,\L_K\Theta_x=\Theta_x
$$
the contraction of the vector field $K$ with 2-form (\ref{GSIIP}), i.e. with the sum of the terms of order $\le 3$ of 2-form (\ref{GSsympl}), is equal to the sum of the terms of order $\le 3$ of the expression
$$
(u +\la w+\la^2 y-\la^3  \Theta_x+\cdots)d(z-\la x-\la^2\Theta_y-\la^3\Theta_w+\cdots),
$$
i.e. to
$$
udz+\la (wdz-udx)+\la^2(ydz-wdx-ud(\Theta_y))+\la^3(-\Theta_xdz-ydx-wd(\Theta_y)-ud(\Theta_w)).
$$
After the restriction to the hypersurface $u=1$ we get
\begin{align}\label{alIIH}\equ
\al^\la|_S=dz+\la (wdz-dx)+\la^2(ydz-wdx-d(\theta_y))+\la^3(-\theta_xdz-ydx-wd(\theta_y)-d(\theta_w))=\nonumber\\
=dz+\la(-dx+wdz) + \la^2(-(\theta_{xy}+w)dx-\theta_{yy}dy+(-\theta_{yz}+y)dz-\theta_{wy}dw)\nonumber\\
\la^3(-(\theta_{wx}+w\theta_{xy}+y)dx-(\theta_{wy}+w \theta_{yy})dy-(\theta_{wz} +w\theta_{yz}+\theta_{x})dz\nonumber\\
-(\theta_{ww}+w\theta_{wy})dw).
\end{align}
Equations (\ref{IIH4D}) are equivalent to the Frobenius integrability condition $(\al^\la|_S)\wedge d(\al^\la|_S)=0$.
\end{rema}

\section{The $f\mapsto \Phi(f)$ symmetry of the Hirota system and ``twisted''  SDVE metrics}
\label{sTwisted}

In this section we will construct a series of examples of SDVE metrics using the $f\mapsto \Phi(f)$ symmetry of the Hirota system, which was discussed in Section \ref{s1}. The scheme of our considerations is as follows.

Firstly let us explain peculiarity of the relations between  $\al^\la|_S$ and $\be^\la|_S=d(\al^\la|_S)$ (note that the degree in $\la$ of a one-form defining a Veronese web in 4D is  greater by one than that of the Gindikin structure) described in Theorems \ref{th1}--\ref{th2} at the common base. To this end consider
a 1-form  $\al^\la=\al_0+\la\al_1+\la^2\al_2+\la^3\al_3$ defining a Veronese web on a manifold of dimension 4 and assume that $d(\al^{\la_0})=0$ for a particular value $\la_0$ of the parameter\footnote{In fact the last condition can be achieved for any $\la_0$ using the integrating multiplier, i.e. a function  $\phi^{\la_0}$ such that $d(\phi^{\la_0}\al^{\la_0})=0$ which exists due to the integrability of $\al^\la$.}.  Now  consider the 2-form given by
 $$
    \be^\la_{\la_0}:=\frac1{\la-\la_0}d\al^\la.
    $$
We claim that it is a polynomial of degree 2 in $\la$.
   Indeed, since $d(\al_0+\la_0\al_1+\cdots+\la_0^{3}\al_3)=0$, we have
    $$
    d\al^\la=d(\al^\la-\al^{\la_0})=(\la-\la_0)d(p(\la,x)),
    $$
where $p(\la,x)$ is a 1-form that is a polynomial of degree 2 in $\la$. Thus $\be^\la_{\la_0}$ has needed degree and, moreover, is obviously closed and satisfies $\be^\la_{\la_0}\wedge \be^\la_{\la_0}\equiv_\la 0$.

In the case of general heavenly equation the situation above is applied to $\la_0=\la_5$ (see formulas (\ref{falpha}) and (\ref{betaGHa}) and notice that $\al^{\la_5}$ coincides with $df$ up to a constant factor). For I and  II heavenly cases we have $\la_0=0$, see formulas (\ref{alIHS}), (\ref{alIIH}).

Secondly, observe that under the assumptions above we have locally $\al^{\la_0}=d\psi$ for some function $\psi$ (the integral) and $\al_\phi^\la:=\phi(\psi)\al^\la$, where $\phi$ is any smooth function of one argument, again satisfies these assumptions. Now ${\be}_\phi^\la:=d\al_\phi^\la$ is again a Gindikin structure, and in general ${\be}_\phi^\la\not=\be^\la$. Below we shall show that the SDVE metrics corresponding to $\be^\la$ and $\be_\phi^\la$ can differ essentially, which opens a way for constructing new examples. We shall refer to metrics corresponding to $\be_\phi^\la$ as twisted (by the function $\phi$).

The connection to $f\mapsto \Phi(f)$ symmetry of the Hirota system is as follows. Formula (\ref{falpha}) for $\al^\la$ shows that if we substitute $f$ by $\Phi(f)$, the 1-form $\al^\la$ will be multiplied by $\phi(f):=\Phi'(f)$.

In the I and II heavenly cases $\la_0=0$ and the integral $\psi$ coincides with $z$,  see formulas (\ref{alIHS}), (\ref{alIIH}).

\abz\label{thTw}
\begin{theo} \begin{enumerate}\item Let a function $\theta(r,s,z,w)$ satisfy system (\ref{IHir4D}). Then the 2-form
$$
\frac1{\la}d(\phi(z)\al^\la),
$$
where $\al^\la$ is given by (\ref{alIHS}) and $\phi$ is any nonzero smooth function of one argument, is a Gindikin structure. The matrix $[g_\phi]^{-1}$ of the inverse to the corresponding SDVE metric $g_\phi$ is given by
$$
[g_\phi]^{-1}=\frac{1}{\phi(\phi'\theta_{ww}+\phi)}\left[
    \begin{array}{cccc}
     0 & -\phi's\theta_{sw} & \phi \theta_{sw} & s\phi'\theta_{ss}-\phi\theta_{sz}\\
      * & 2\phi's\theta_{rw} & -\phi\theta_{rw}  & \phi\,\theta_{rz}+\phi'(\theta_r-s\theta_{rs}) \\
      * & * & 0 & 0 \\
      * & * & * & 2{\phi'}
    \end{array}
  \right].
$$
\item Let a function $\theta(x,y,z,w)$ satisfy system (\ref{IIH4D}). Then the 2-form
$$
\frac1{\la}d(\phi(z)\al^\la),
$$
where $\al^\la$ is given by (\ref{alIIH})  and $\phi$ is any nonzero smooth function of one argument, is a Gindikin structure. The matrix $[g_\phi]^{-1}$ of the inverse to the corresponding SDVE metric $g_\phi$ is given by
$$
[g_\phi]^{-1}=\frac{1}{\phi'\theta_{yy}+\phi}\left[
    \begin{array}{cccc}
     -2 \theta_{yy}  & 2\theta_{xy}-\frac{\phi'}{\phi}\theta_{wy} & 0 & \frac{\phi'}{\phi}\theta_{yy}-1 \\
      * & 2(-\theta_{xx}+\frac{\phi'}{\phi}(\theta_{wx}+y)) & -1  & -\frac{\phi'}{\phi}(\theta_{xy}-w) \\
      * & * & 0 & 0 \\
      * & * & * & 2\frac{\phi'}{\phi}
    \end{array}
  \right].
$$
 \end{enumerate}
\end{theo}

\noindent{\sc Proof} The proof of the first statements of Item 1,2 follows from the considerations above and from Theorems \ref{th1a}, \ref{th2}. For the direct calculation of the metric (which we skip) we use the Gindikin formula for restoring the matrix $[g]$ of the SDVE metric $g$ from its Gindikin form $\be^\la=\be_0+\la\be_1+\la^2\be_2$ \cite{gindikin82}:
$$
[g]=-\frac{1}{\mu_0\nu_1-\mu_1\nu_0}[\overline{\be}^{\mu+\nu}-\overline{\be}^\mu-
\overline{\be}^\nu][\overline{\be}^\mu+\overline{\be}^\nu]^{-1}[\overline{\be}^\mu-\overline{\be}^\nu],
$$
where $\mu=(\mu_0,\mu_1)$, $\nu=(\nu_0,\nu_1)$, $\overline{\be}^\mu=\mu_0^2\be_0+\mu_0\mu_1\be_1+\mu_1^2\be_2$ is the ``homogenization'' of the polynomial $\be^\la$, and $[\cdot]$ in the right hand side is the matrix of the corresponding 2-form. It turns out that the matrix $[g]$ of the metric $g$ is more complicated than $[g]^{-1}$, that is why we present the formula for the inverse matrix. \qed

\abz\label{remm}
\begin{rema}\rm
It is worth to mention that the matrices of the inverse to  ``untwisted'' metrics (\ref{metrIH}) and (\ref{metrIIH})
$$
\left[
  \begin{array}{cccc}
   0   & 0& \theta_{sw} & -\theta_{sz}  \\
    * & 0 & -\theta_{rw} & \theta_{rz}   \\
    * & * & 0 & 0  \\
    * & * & * & 0 \\
  \end{array}
\right],
\left[
  \begin{array}{cccc}
   -2 \theta_{yy}  & 2 \theta_{xy}& 0 & -1  \\
    * & -2 \theta_{xx} & -1 & 0  \\
    * & * & 0 & 0  \\
    * & * & * & 0 \\
  \end{array}
\right]
$$
are obtained by putting $\phi=1$ in the formulas of the theorem.
\end{rema}

\abz
\begin{exa}\rm
Take the trivial solution $\theta=0$ of the II heavenly PDE corresponding to the flat metric. Then  ``twisting'' by any smooth function $\phi$ gives the metric $g_\phi=-2 \phi'dx^2-w \phi'dx\ dz-\phi dx\ dw-\phi dy\ dz - 2 y\phi'dz^2$ with the Weyl spinor
$$
\frac{\phi'''}{\phi^2}D_{X^0}^4.
$$
This effect, that the symmetry $f\mapsto\Phi(f)$ can produce from the flat metric a nonflat one, was observed already in \cite{konopSchiefszer}.

\end{exa}

\abz\label{ppw}
\begin{exa}\rm
Take an obvious solution $\theta(x,y,z,w)=F(y,w)$ of the II heavenly PDE (the so called $pp$-waves). Then the corresponding SDVE metric $g=dwdx+dydz-F_{yy}dw^2$ is nonflat with the Weyl spinor
$$
C=8F_{yyyy}D_{X^1}^4
$$
and fundamental invariants $I = \tensor{C}{_{ABCD}}\tensor{C}{^{ABCD}}=0$ and $J=\tensor{C}{_{AB}^{CD}}\tensor{C}{_{CD}^{EF}}\tensor{C}{_{EF}^{AB}}=0$ (cf. \cite{dunajskiWest07,penroseRindler}).

It is easy to see that the function $\theta(x,y,z,w)=F(y,w)$ satisfies system (\ref{IIH4D}) (the full Hirota system for II Plebański equation) if and only if
\begin{equation}\label{lapl}\equ
 yF_{yy}+wF_{wy}+F_{ww}=0.
\end{equation}
Take a particular solution $F(y,w)=(4y-w^2)^{3/2}$. Then after ``twisting''
\begin{itemize}\item for $\phi(z)=z$ we have $I\not=0$, $J\not=0$, $I^3-6J^2=0$;
\item for $\phi(z)=z^2$ we have $I=0$, $J=0$, $I^3-6J^2=0$;
\item for $\phi(z)=z^3$ we have $I\not=0$, $J\not=0$, $I^3-6J^2\not=0$.
\end{itemize}
This example shows that starting from an algebraically very special solution of the II heavenly PDE the ``twisting'' trick can produce even an algebraically general solution. We refer the reader to Appendix \ref{apB}, where  the general solution of equation (\ref{lapl}) is found and the case of the above solution is elaborated in detail (including explicit form of the metric, its Weyl spinor, and its fundamental invariants).
\end{exa}

\abz
\begin{rema}\rm
It is worth mentioning that equation (\ref{lapl}) appeared in \cite[Sect. 2.1]{dunajski} in a different context. The solutions of this equation describe a subclass of Lorentzian hyper-CR Einstein--Weyl structures.
\end{rema}

\abz
\begin{exa}\rm
One can check that the function
$$
\theta:=(c_1(r+z)+c_2w)s-
\frac1{c_1^2}\left({c_2c_3}e^{-\frac{c_1}{c_2}(c_1(r+z)+c_2w)}-(c_1w-1)(r+z)+\frac{c_1c_2w^3}{6}
-\frac{c_2w^2}{2}\right),
$$
where $c_i$ are arbitrary nonzero constants, satisfies system (\ref{IHir4D}) corresponding to the Hirota system in the framework of I Plebański equation and that the corresponding ``untwisted'' metric is nonflat. We are able to perform the twisting procedure using one-form (\ref{alIHS}). To simplify cumbersome expressions, we put $c_1=c_2=c_3=1$. The inverse twisted metric is given by
$$
[g_\phi]^{-1}=\frac1{\phi(\phi'(-E-w+1)+\phi)}\left[
    \begin{array}{cccc}
     0 & -\phi's & \phi  & -\phi\\
      * & -2\phi's(E-1) & -\phi (E-1)  &  (-\phi+\phi')E+\phi'(w-1) \\
      * & * & 0 & 0 \\
      * & * & * & 2{\phi'}
    \end{array}
  \right],
$$
where we write $E:=e^{-r-z-w}$.
The Weyl spinor for the corresponding twisted metric $g_\phi$ is given by
\begin{align*}
C=\frac{\phi E((\phi'(2E-w+1)+\phi)}{(\phi'(-E-w+1)+\phi)^5}D^4_{X^0}-4\frac{\phi' E((\phi'(2E-w-2)+\phi)}{(\phi'(-E-w+1)+\phi)^4}D^3_{X^0}D_{X^1}\\
+6\frac{2(\phi')^3 E^2-2(\phi')^2\phi E+\phi^2\phi''E-4(\phi')^3E+2(\phi')^3}{(\phi'(-E-w+1)+\phi)^3\phi}D^2_{X^0}D^2_{X^1}\\
-\frac4{(\phi'(-E-w+1)+\phi)^2\phi^2}(2(\phi')^3 E^2+2(\phi')^3Ew-8(\phi')^2\phi E+3\phi^2\phi''E-4(\phi')^3E\\
-2(\phi')^3w+8(\phi')^2\phi-
3\phi^2\phi''+2(\phi')^3)\phi'D_{X^0}D^3_{X^1}\\
+\frac1{(\phi'(-E-w+1)+\phi)\phi^3}(2(\phi')^5E^2+4(\phi')^5Ew-10(\phi')^4\phi E-\phi'''\phi'\phi^3E
+3(\phi'')^2\phi^3E-4(\phi')^5E\\
+2(\phi')^5w^2-10 (\phi')^4\phi w-\phi'''\phi'\phi^3w+3(\phi'')^2\phi^3w-4(\phi')^5w+20(\phi')^3\phi^2-12\phi'\phi^3\phi''\\
+\phi'''\phi^4+10(\phi')^4\phi+\phi'''\phi'\phi^3-3(\phi'')^2\phi^3+2(\phi')^5)D_{X^1}^4.
\end{align*}
For $\phi=1$, i.e. for the ``untwisted'' metric, it is equal to $ED^4_{X^0}$ with trivial fudamental invariants, however already for $\phi=z$ we get $I\not=0$, $J\not=0$ and $I^3-6 J^2\not=0$.

\end{exa}

\section{Uniqueness of the Hirota system}
\label{suniq}

In this section we address the question of uniqueness of the Hirota system, which in our approach is reduced to the following matter: given a Gindikin structure $\be^\la$ with a symmetry $K$ with the constant $c\not=0$, see Definition \ref{symm}, what is the most general form of the pair $(\be^\la,K)$? We try to answer to this question in the framework of general heavenly equation. To this end we shall exploit the notion of ``privileged'' coordinates, which was used in \cite{konopSchiefszer} in 4D.

\abz\label{adap}
\begin{defi}\rm
Let $\be^\la$ be a Gindikin structure in 5D and let $X_1^\la,X_2^\la$, and $X_3^\la$ be three linearly independent vector fields spanning the integrable distribution $\ker\be^\la$. Fix $\la_1,\ldots,\la_5$, 5 pairwise different values of $\la$. We say that coordinates $(\phi^1,\ldots,\phi^5)$ are adapted to $(\la_1,\ldots,\la_5)$ if $X_j^{\la_i}\iprod d\phi^i=0$ (no summation) for any $j=1,2,3$ and any $i=1,\ldots,5$ (equivalently, $d\phi^i\wedge \be^{\la_i}=0$, $i=1,\ldots,5$).
\end{defi}
In general, since $\ker\be^\la$ is of codimension 2, there is an ambiguity of two functional dimensions in the choice of each of the adapted coordinates for a fixed 5-tuple $\boldsymbol{\la}=(\la_1,\ldots,\la_5)$.

Coming back to the special case of the general heavenly framework we remark that the coordinates $(x^1,\ldots,x^5)$ are adapted to $\boldsymbol{\la}=(\la_1,\ldots,\la_5)$ in the case of Gindikin structure (\ref{betaGH}) in 5D. It turns out that, moreover, $(F_1(x^1,\frac{\partial f }{\partial x^1}),\ldots,F_5(x^5,\frac{\partial f }{\partial x^5}))$ are adapted to $\boldsymbol{\la}$ for any functions $F_i$ of two arguments (provided the functional independence).

 Moreover,  under the assumption of Item 2 of Theorem \ref{th1}, putting $\tilde{q}(x^5)=\ln(q(x^5))$, we get $K=\frac{q(x^5)}{q'(x^5)}\frac{\partial}{\partial x^5}=\frac{\partial\ }{\partial \tilde{q}}$. The new coordinates $(x^1,\ldots,x^4,\tilde{q}(x^5))$
are also adapted to $\boldsymbol{\la}$. Corollary \ref{corO} of the following two theorems says that in fact any Gindikin structure with a symmetry $K$ with the constant $c\not=0$ in 5D can be put in  form (\ref{betaGH}) and $K$ can be put in the form as above.

\abz\label{EP}
\begin{theo}
Let $\be^\la$ be a Gindikin structure in 5D. Fix $\la_1,\ldots,\la_5$,  pairwise different values of $\la$. Then for any coordinates $(x^1,\ldots,x^5)$ adapted to $\boldsymbol{\la}=(\la_1,\ldots,\la_5)$  the structure $\be^\la$ is given by formula (\ref{betaGH}) for an appropriate smooth function $f$.
\end{theo}

Before starting the proof we shall make some introductory remarks and prove a lemma. The proof will follow the lines of the proof of the existence of the potential performed in \cite{pHeavenly} in 4D for a quite general class of PDEs describing the SDVE metrics.

The idea of the proof is as follows. We shall use the Nijenhuis operator $\widetilde{N}:TM\to TM$ given by formula (\ref{NOp}) of Appendix A. One associates the so-called Nijenhuis differential $d_{\widetilde{N}}$ with such an operator\footnote{\label{ftnO} It is given by $d_N=[i_N,d]$, where
$
(i_{\widetilde{N}}\al)(X_1,X_2,\ldots,X_k)=\al(\widetilde{N}X_1,X_2\ldots,X_k)+
\al(X_1,\widetilde{N}X_2\ldots,X_k)+\cdots+\al(X_1,X_2\ldots,\widetilde{N}
X_k),
$
$\al$ is a differential $k$-form on $M$.} and we shall prove that the 2-form $\be^\infty$ id $d_{\widetilde{N}}$-closed. Since it is also $d$-closed we then use the corresponding ``$dd_{\widetilde{N}}$-lemma'' for proving the existence of a function $f$ such that $\be^\infty=dd_{\widetilde{N}}f$. At last we show that $\be^\la$ coincides up to a constant factor with $dd_{(\widetilde{N}-\la \Id)^{-1}}f$, which is the needed form of $\be^\la$.

We refer the reader to Appendix A, where we show that the kernel $\ker\be^\la$ is tangent to a generic Kronecker web of codimension 2 that is divergence free with respect to some volume form $\om$. This in particular means the existence of linearly independent vector fields $X_1,X_2,X_3$ spanning $\ker\be^\infty$ and such that the vector fields $X_j^\la:=(\widetilde{N}-\la \Id)X_j$, $j=1,2,3$, span $\ker\be^\la$, pairwise commute, and are divergence free with respect to $\om$. We now are able to prove  the following lemma.

\abz\label{lemk}
\begin{lemm}
Retain the assumption of Theorem \ref{EP} and the notations above. Then the two-form $\be^\infty=\iota(X_1\wedge X_2\wedge X_3)\om$ (cf. Remark \ref{remAp}) is $d_{\widetilde{N}}$-closed.
\end{lemm}

\noindent{\sc Proof}  This lemma is a generalization of \cite[Th. 3.4(ii)]{pHeavenly}. Note that $\Tr(\widetilde{ N})$ is constant, thus condition (3.4) of \cite[Th. 3.4]{pHeavenly} is satisfied and we can extend the proof of this theorem.

$d_{\widetilde{N}}$-closedness of the form $\be^\infty=\iota(X_1\wedge X_2\wedge X_3)\om$ is equivalent to $D_{\widetilde{N}}$-closedness of the trivector $X_1\wedge X_2\wedge X_3$. Here $D_{\widetilde{N}}$ is the Koszul--Nijenhuis operator \cite{pHeavenly} given by
$$
D_N=\Phi^{-1}\circ d_N\circ \Phi:\chi^k(M)\to\chi^{k-1}(M),
$$
where $\chi^k(M)$ stands for the space of $k$-vector fields on $M$ and $\Phi$ is the isomorphism of  $\chi^k(M)$ with the space of $(5-k)$-forms given by the contraction with $\om$. Now we shall prove the $D_{\widetilde{N}}$-closedness of the trivector $X_1\wedge X_2\wedge X_3$.

We shall use the following formulas:
\begin{itemize}\item
\begin{equation}\label{eDN}\equ
[X_i,X_j]=0,[\widetilde{N}X_i,\widetilde{N}X_j]=0,[\widetilde{N}X_i,X_j]+[X_i,\widetilde{N}X_j]=0,i,j=1,2,3,
\end{equation}
 which follow from the commutation relations $[X_i^{\la},X_j^\la]=0$;
 \item formula \cite[form. (3.2)]{pHeavenly}
\begin{equation}\label{eDN}\equ
D_{\widetilde{N}}(U)=-[j_{\widetilde{N}},D](U)-[\Tr(\widetilde{N}),U]_S=-[j_{\widetilde{N}},D](U), \qquad U\in\chi^k(M),
\end{equation}
where the term with the Schouten bracket $[,]_S$ vanishes since the trace $\Tr(\widetilde{N})$ is constant, $j_{\widetilde{N}}$ is given by
$$
j_{\widetilde{N}}(X_1\wedge X_2\wedge\cdots \wedge X_k):=\widetilde{N}X_1\wedge X_2\wedge\cdots \wedge X_k+X_1\wedge \widetilde{N}X_2\wedge\cdots\wedge X_k+\cdots+X_1\wedge X_2\wedge\cdots \wedge \widetilde{N}X_k
$$
and $D=\Phi^{-1}\circ d\circ \Phi:\chi^k(M)\to\chi^{k-1}(M)$ is the Koszul differential, and
 \item the Koszul formula \cite{koszul}
\begin{equation}\label{DKos}\equ
[U,V]_S=(-1)^k (D(U\wedge V)-D(U)\wedge V-(-1)^k U\wedge D(V)), \qquad  U\in\chi^k(M), \quad V\in\chi^l(M).
\end{equation}
 \end{itemize}

Since $d\be^\infty=d(\iota(X_1\wedge X_2\wedge X_3)\om)$, we get $D(X_1\wedge X_2\wedge X_3)=0$ and, due to the fact that $X_i^\la$ are divergence free, also $D(X_i)=0$ and $D(\widetilde{N}X_i)=0$, $i=1,2,3$. Thus
\begin{align*}
D_{\widetilde{N}}(X_1\wedge X_2\wedge X_3)=-[j_{\widetilde{N}},D](X_1\wedge X_2\wedge X_3)\\
=D(\widetilde{N}X_1\wedge X_2\wedge X_3+X_1\wedge \widetilde{N}X_2\wedge X_3+X_1\wedge X_2\wedge \widetilde{N}X_3)\\
=D(\widetilde{N}X_1\wedge X_2)\wedge X_3+(\widetilde{N}X_1\wedge X_2)\wedge D(X_3)+[\widetilde{N}X_1\wedge X_2,X_3]_S\\
+D(X_1\wedge \widetilde{N}X_2)\wedge X_3+(X_1\wedge \widetilde{N}X_2)\wedge D(X_3)+[X_1\wedge \widetilde{N}X_2,X_3]_S\\
+D(X_1\wedge X_2)\wedge \widetilde{N}X_3+(X_1\wedge X_2)\wedge D(\widetilde{N}X_3)+[X_1\wedge X_2,\widetilde{N}X_3]_S\\
=(-[\widetilde{N}X_1,X_2]-D(\widetilde{N}X_1)\wedge X_2+\widetilde{N}X_1\wedge D(X_2))\wedge X_3+[\widetilde{N}X_1\wedge X_2,X_3]_S\\
(-[X_1,\widetilde{N}X_2]-D(X_1)\wedge \widetilde{N}X_2+X_1\wedge D(\widetilde{N}X_2))\wedge X_3+[X_1\wedge \widetilde{N}X_2,X_3]_S\\
(-[X_1,X_2]-D(X_1)\wedge X_2+X_1\wedge D(X_2))\wedge \widetilde{N}X_3+[X_1\wedge X_2,\widetilde{N}X_3]_S\\
=[\widetilde{N}X_1,X_3]\wedge X_2+\widetilde{N}X_1\wedge[X_2, X_3]+[X_1,X_3]\wedge \widetilde{N}X_2+X_1\wedge[\widetilde{N}X_2,X_3]\\
+[X_1,\widetilde{N}X_3]\wedge X_2+X_1\wedge [X_2,\widetilde{N}X_3]\\
=([\widetilde{N}X_1,X_3]+[X_1,\widetilde{N}X_3])\wedge X_2+X_1\wedge([\widetilde{N}X_2,X_3]+[X_2,\widetilde{N}X_3])=0.
\end{align*}
\qed

\noindent{\sc Proof of Theorem \ref{EP}} Once we have $d$- and $d_{\widetilde{N}}$-closedness of the 2-form $\be^\infty$ (the last established in Lemma \ref{lemk}), we can use the ``$dd_{\widetilde{N}}$-lemma'' saying that, if a Nijenhuis operator ${\widetilde{N}}$ is cyclic at any point (our diagonal ${\widetilde{N}}$ given by (\ref{NOp}) is so), then any 2-form which is $d$- and $d_{\widetilde{N}}$-closed is equal to $dd_{\widetilde{N}}f$ for some function $f$, see \cite{turielDDN,bolsKonMat},\cite[Th. 3.1]{pHeavenly}.

Consider the two-form
$$
\zeta^\la:=(\la-\la_1)\cdots(\la-\la_{5})dd_{(\widetilde{N}-\la \Id)^{-1}}f,
$$
where $f$ is such that $\be^\infty =dd_{\widetilde{N}}f$.
One checks that this form coincides with that given by (\ref{betaGH})  (cf. footnote \ref{ftnO} on page \pageref{ftnO}). By \cite[Lemma 4.3]{pHeavenly} we conclude that the forms $\be^\la$ and $\zeta^\la$ have the same kernel $({\widetilde{N}}-\la\Id)D^\infty$, where $D^\infty=\ker\be^\infty$. On the other hand, it follows from the proof of Theorem \ref{1to1} that they have to differ by a nonvanishing factor not depending on $\la$: $\beta^\la=a\zeta^\la$. Finally, since $d\beta^\la=0$, the function $a$ has to be constant and can be adsorbed by $f$. \qed

\abz
\begin{theo}\label{thuniq}
Let $\be^\la$ be a Gindikin structure in 5D and let $K$ be its symmetry with the constant $c\not=0$. Fix $\la_1,\ldots,\la_5$, pairwise different values of $\la$. Then away from the singularities of $K$ there exist adapted to $\boldsymbol{\la}=(\la_1,\ldots,\la_5)$ coordinates $(x^1,\ldots,x^5)$ such that $K=\frac{\partial\ }{\partial x^5}$. The coordinates $(x^1,\ldots,x^4)$ are defined uniquely up to a change $x^i\mapsto \phi_i(x^i)$ (no summation), where $\phi_i$, $i=1,\ldots,4$, are functions of one argument. The coordinate $x^5$ is defined uniquely up to the addition of any function $g=g(x^1,\ldots,x^4)$ such that $dg\wedge\be^{\la_5}=0$.
\end{theo}

\noindent{\sc Proof} Not restricting generality we may assume that $c=1$. We notice that $K$ does not belong to $\ker\be^\la$ for any $\la$. Otherwise, if $K\in\ker\be^{\la_0}$ for some $\la_0$, then $\L_K\be^{\la_0}=d(K\iprod \be^{\la_0})=0$, which is a contradiction. On the other hand,
\begin{equation}\label{ideal}\equ
\iota_{[K,X_j^{\la_i}]}\be^{\la_i}=[\L_K,\iota_{X_j^{\la_i}}]\be^{\la_i}=-\iota_{X_j^{\la_i}}\L_K\be^{\la_i}=
-\iota_{X_j^{\la_i}}\be^{\la_i}=0,
\end{equation}
in particular, $D_i:=\Span(K,X_1^{\la_i},X_2^{\la_i},X_3^{\la_i})$ is an integrable distribution for any $i=1,\ldots,5$. From this we conclude that there exist coordinates $(x^1,\ldots,x^5)$ adapted to $\boldsymbol{\la}$ such that $K=\frac{\partial\ }{\partial x^5}$. Indeed, the coordinates $x^1,\ldots,x^4$ are defined by the condition $dx^i(D_i)=0$, $i=1,\ldots,4$, with the ambiguity $x^i\mapsto \phi_i(x^i)$. The coordinate $x^5$ is chosen as follows. Once $x^1,\ldots,x^4$ are built,  choose $\phi^1$ to be any coordinate such that $(x^1,\ldots,x^4,\phi^1)$ are adapted to $\boldsymbol{\la}$. Then $K$ takes the form $K=F(\phi^1,\phi^2)\frac{\partial\ }{\partial \phi^1}$,  where $F$ is some function of two arguments and $\phi^2$ is the second coordinate defined by the condition $d\phi^2(X_j^{\la_5})=0$, $j=1,2,3$, and additionally by $d\phi^2(D_5)=0$ (note that de facto $\phi^2$ is functionally dependent on $x^1,\ldots,x^4$). This form of $K$ follows from the  fact that $[X_j^{\la_5},K]\in\ker\be^{\la_5}$, $j=1,2,3$, as (\ref{ideal}) shows, and that $K$ does not belong to $\ker\be^{\la_5}$. Finally, putting $x^5:=\int (1/F(\phi^1,\phi^2))d\phi^1$ we get that $(x^1,\ldots,x^5)$ are adapted to $\boldsymbol{\la}$ and $K=\frac{\partial\ }{\partial x^5}$. \qed

\abz\label{corO}
\begin{coro}
Let $\be^\la$ be a Gindikin structure with a symmetry $K$ with the constant $c\not=0$ (see Definition \ref{symm}) in 5D. Fix $\la_1,\ldots,\la_5$,  pairwise different values of $\la$. Then away from the singularities of $K$ there exist adapted to $(\la_1,\ldots,\la_5)$ coordinates $(x^1,\ldots,x^5)$ such that $K=\frac{\partial\ }{\partial x^5}$ and $\be^\la$ is given by formula (\ref{betaGH}), where the function  $f$ is of the form $f=g(x^1,\ldots,x^4)\cdot e^{x^5}$ with $g$, a smooth function of four variables.
\end{coro}

\noindent{\sc Proof} We only have to prove that $f$ has to attain the appropriate form. Indeed, put $c=1$. One checks that if $\be^\la$ is of the form (\ref{betaGH}) and $K=\frac{\partial\ }{\partial x^5}$, the function $f$ has to satisfy the following system of PDEs implied by the condition $\L_K\be^\la=\be^\la$:
$$
f_{ij}=f_{ij5},1\le i<j\le 5.
$$
Its general solution is $f(x)=g(x^1,\ldots,x^4)\cdot e^{x^5}+\sum_{i=1}^5h_i(x^i)$, where $h_i$ are some functions of one variable. The last term can be neglected as it does not contribute to $\be^\la$. \qed

\section{Concluding remarks}
\label{sConcl}

We would like to accent on some perspectives.
\begin{itemize}\item It was shown in \cite[Th. 10.1]{konopSchiefszer} that a generic SDVE metric coming form the dispersionless Hirota system does not admit conformal Killing vectors. It seems that in order to find some explicit cases of such a metrics one may try to start from some solutions admitting Killing vectors and then apply the twisting (see Section \ref{sTwisted}) in order to reduce a number of them.
\item We have introduced some symmetry reduction procedure of 5-dimensional Gindikin structures leading to a subclass of 4-dimensional Gindikin structures related to Veronese webs in 4D, or equivalently to solutions of the Hirota dispersionless system. This reduction is depicted as the left column of the following diagram:
$$
\begin{array}{ccc}
\tilde\be^\la\ \mbox{(5D\ Gindikin\ structure)}  & \longleftrightarrow & \tilde g\ \mbox{(unknown geometric structure)} \\
\downarrow  /K & & \downarrow  /K \\
\be^\la\ \mbox{(4D\ Gindikin\ structure)} &\longleftrightarrow & g\ \mbox{(SDVE metric)} \;.
\end{array}
$$
Since 4-dimensional Gindikin structures correspond to SDVE metrics (the lower horizontal row) we get a subclass of these last. A natural question arises, is it some natural $\la$-independent geometric structure $\tilde g$ corresponding to a 5D Gindikin structure ($\tilde\be^\la$ and $\tilde g$ would obey the same symmetries similarly to $\be^\la$ and $g$, see the discussion after Theorem \ref{gii})\footnote{In a private discussion Prof. Maciej Dunajski suggested that this could be a specific symmetric 3-form.}.  The condition of closedness for $\tilde\be^\la$ would correspond to a kind of ``half-flatness'' of $\tilde g$ and the coefficients of $\tilde\be^\la$ in $\la$ should serve as a generalization  of the self-dual forms of $g$. Finally, recall that our definition of the symmetry $K$ of $\tilde\be^\la$ (see Definition \ref{symm}) is based on the notion of a triholomorphic symmetry of $g$, which preserves each self-dual form individually. However, there exist more general symmetries of $g$ which preserve only the space of self-dual forms \cite{dunajskiTod}. We hope that finding the structure $\tilde g$ and studying its general symmetries would give new insights in the investigation of SDVE metrics.
\end{itemize}

\appendix

\section{Appendix: Gindikin structures and divergence free Kronecker webs in 5D}
\label{apA}

The notion of a divergence free Kronecker webs was introduced in \cite{pHeavenly}, where also a one-to-one correspondence between divergence free Kronecker webs of a special type and Gindikin structures was established in 4D. The aim of this appendix is to prove analogous result in 5D (it is used in the proof of Theorem \ref{EP}).

We refer the reader to \cite{z1,pHeavenly} for the definition of a Kronecker web. Recall that this is a collection of foliations $\{\F_\la\}_{\la\in\R\P^1}$ on a manifold $M$ such that there exist a partial Nijenhuis operator $N:T\F_\infty\to TM$ such that $(N-\la I)T\F_\infty=T\F_\la$ for any $\la\in\R\P^1$, where $I:T\F_\infty\to TM$ is the canonical inclusion.
One can understand $N$ as a restriction to $T\F_\infty$ of a not necessarily unique Nijenhuis operator $\widetilde{N}:TM\to TM$ such that $\widetilde{N}(T\F_\infty)$ is an integrable distribution, see \cite[Lemma 1.2]{pHeavenly}.
We will need the following specialization a notion of a Kronecker web.

\abz
\begin{defi}\rm We say that a Kronecker web  $\{\F_\la\}$, $T\F_\la=(N-\la I)T\F_\infty$, of codimension two on a 5-dimensional manifold $M$ is  \emph{generic} if the Jordan--Kronecker decomposition of the pair of operators $N_x,I_x:T_x\F_\infty\to T_xM$ for any $x\in M$ consists of two Kronecker blocks whose matrices have dimensions $3\times 2$ and $2\times 1$ respectively. Equivalently, a collection $\{\F_\la\}$ of foliations of codimension two is a generic Kronecker web of codimension two if
locally there exist linearly independent vector fields $Y_1,\ldots,Y_5$ such that
$$
T\F_\la=\Span(\la Y_1- Y_2,\la Y_2- Y_3,\la Y_4-Y_5).
$$
\end{defi}

The terminology is motivated by the fact that  such a situation is generic (in comparison to more degenerate situation with two Kronecker blocks with matrices of dimensions $4\times 3$ and ``$1\times 0$'' or more than two Kronecker blocks). A generic Kronecker web of codimension two is annihilated by the two-form $\be^\la=(\ga_0+\la\ga_1+\la^2\ga_2)\wedge(\de_0+\la\de_1)$, where $\ga_0,\ldots,\de_1$ is the local coframe dual to the local frame $Y_1,\ldots,Y_5$, i.e. by a ``nearly-Gindikin structure'' (instead of the closedness of $\be^\la$, the weaker  condition of the complete integrability of the differential system generated by $\ga^\la:=\ga_0+\la\ga_1+\la^2\ga_2$ and $\de^\la:=\de_0+\la\de_1$ is required). The following notion specifies a subclass of webs corresponding to Gindikin structures, as Theorem \ref{1to1} shows.

\abz\label{divfreedefi}
\begin{defi}\rm\cite{pHeavenly}
Let $\om$ be a volume form on a 5-dimensional manifold $M$. A generic Kronecker web $\{\F_\la\}_{\la\in\R\P^1}$,  $T\F_\la=(N-\la I)T\F_\infty$, of codimension two on  $M$ is said to be  \emph{divergence free} with respect to $\om$ if there exist  vector fields $X_1,X_2,X_3$    such that for any $\la\in\R\P^1$ the vector fields $X_1^\la,X_2^\la,X_3^\la$,\, $X_i^\la:=(N-\la I)(X_i)$,
\begin{enumerate}\item  are linearly independent and span $T\F_\la$;
\item  pairwise commute: $[X_i^\la,X_j^\la]=0$;
\item  are divergence free with respect to $\om$: $d(X_i^\la\iprod\om)=0$.
\end{enumerate}
\end{defi}

\abz\label{1to1}
\begin{theo}
Let $\be^\la$ be a Gindikin structure on a 5-dimensional manifold $M$. Then there exist a volume form $\om$ on $M$ such that the integrable distribution $\ker\be^\la$ is tangent to a generic Kronecker web that is divergence free with respect to $\om$.

Vice versa, given a generic Kronecker web $\{\F_\la\}$ of codimension two on $M$ that is divergence free with respect to a volume form $\om$, the 2-form $\be^\la=X_1^\la\wedge X_2^\la\wedge X_3^\la\iprod \om$, where $X_i^\la$ are the vector fields from Definition \ref{divfreedefi}, is a Gindikin structure on $M$.
\end{theo}

\noindent{\sc Proof}
Let  $\be^\la$ be a Gindikin structure. We already know that its kernel $\ker\be^\la$ is tangent to a generic Kronecker web $\{\F_\la\}$ of codimension two, i.e. there exists a partial Nijenhuis operator $N:T\F_\infty\to TM$ such that $(N-\la I)T\F_\infty=T\F_\la$ for any $\la\in\R\P^1$.

Fix $\la_1,\ldots,\la_5$,  pairwise different values of $\la$ and let $(x^1,\ldots,x^5)$ be coordinates adapted to $(\la_1,\ldots,\la_5)$, see Definition \ref{adap}.
We claim that the Nijenhuis operator
\begin{equation}\label{NOp}\equ
\widetilde{N}:\partial_{x^j}\mapsto \la_j\partial_{x^j}, \qquad j=1,\ldots,5,
\end{equation}
is an extension of $N$. Indeed, we have
\begin{equation}\label{eX12}\equ
X_i^{\la_j}x^j=((N-\la_j I)X_i)x^j=0,
\end{equation}
(no summation) for any $i=1,2,3$ and any $j=1,\ldots,5$.
The fact that $\widetilde{N}$  is an extension of the partial Nijenhuis operator $N:X_i\mapsto Y_i:=NX_i$ now is obvious: if $X_i=X_i^j\partial_{x^j}$, $Y_i=Y_i^j\partial_{x^j}$, $i=1,2,3$, then (\ref{eX12}) implies $Y_i^j=\la_jX_i^j$ (no summation) and $\widetilde{N}X_i=Y_i$.

Consider any three linearly independent vector fields spanning the integrable distribution $T\F_\infty$. By the standard procedure, after possible permutation of variables and taking appropriate linear combination we obtain the following pairwise commuting vector fields:
\begin{align*}
X_1:=X_{11}\partial_{x^1}+X_{12}\partial_{x^2}+\partial_{x_3},
X_2:=X_{21}\partial_{x^1}+X_{22}\partial_{x^2}+\partial_{x_4},
X_3:=X_{31}\partial_{x^1}+X_{32}\partial_{x^2}+\partial_{x_5}
\end{align*}
with some functions $X_{ij}$. One checks that any 2-form that is a polynomial of third degree in $\la$ and annihilates the vector fields
\begin{align}\label{X`i}\equ
X_1^\la:=(N-\la I)X_1=X_{11}(\la_1-\la)\partial_{x^1}+X_{12}(\la_2-\la)\partial_{x^2}+
(\la_3-\la)\partial_{x_{3}}\nonumber\\
X_2^\la:=(N-\la I)X_2=X_{21}(\la_1-\la)\partial_{x^1}+X_{22}(\la_2-\la)\partial_{x^2}+
(\la_4-\la)\partial_{x_{4}}\nonumber\\
X_3^\la:=(N-\la I)X_3=X_{31}(\la_1-\la)\partial_{x^1}+X_{32}(\la_2-\la)\partial_{x^2}+
(\la_5-\la)\partial_{x_{5}}
\end{align}
is given by
\begin{align*}
\be^\la=a(\la-\la_1)\cdots(\la-\la_{5})\left(\frac{dx^1\wedge dx^2}{(\la-\la_1)(\la-\la_2)}-\frac{X_{12}dx^1\wedge dx^3}{(\la-\la_1)(\la-\la_3)}-\frac{X_{22}dx^1\wedge dx^4}{(\la-\la_1)(\la-\la_4)}\right.\\
-\frac{X_{32}dx^1\wedge dx^5}{(\la-\la_1)(\la-\la_5)}+\frac{X_{11}dx^2\wedge dx^3}{(\la-\la_2)(\la-\la_3)}+\frac{X_{21}dx^2\wedge dx^4}{(\la-\la_2)(\la-\la_4)}
+\frac{X_{31}dx^2\wedge dx^5}{(\la-\la_2)(\la-\la_5)}\\
\left.+\frac{M_{3}dx^3\wedge dx^4}{(\la-\la_3)(\la-\la_4)}+\frac{M_{2}dx^3\wedge dx^5}{(\la-\la_3)(\la-\la_5)}+\frac{M_{1}dx^4\wedge dx^5}{(\la-\la_4)(\la-\la_5)}
\right),
\end{align*}
where $a$ is a function independent of $\la$ and $M_i$, $i=1,2,3$, are the minors of the matrix
$$
\left[
  \begin{array}{ccc}
   X_{11} & X_{21} & X_{31} \\
    X_{12} & X_{22} & X_{32} \\
  \end{array}
\right]
$$
corresponding to deleting the $i$-th column.
In particular, the initial Gindikin structure is of this form with a nonvanishing $a$ and we can put
\begin{equation}\label{vf}\equ
\om:=\frac{1}{(\la-\la_3)(\la-\la_4)(\la-\la_5)}\be^\la\wedge dx^3\wedge dx^4\wedge dx^5=a\,dx^1\wedge\cdots\wedge dx^5.
\end{equation}
Since $X_i^\la\iprod \be^\la=0$ and $X_i^\la\iprod dx^{i+2}=\la_{i+2}-\la$, $i=1,2,3$, the vector fields $X_i^\la$ are divergence free with respect to $\om$.

To prove the second part of the theorem use \cite[Th. 2.2]{pHeavenly} to deduce the closedness of the form $\be^\la:=X_1^\la\wedge X_2^\la\wedge X_3^\la\iprod \om$. The genericity of the Kronecker web will assure the nondegeneracy of the Gindikin structure. \qed

\abz\label{remAp}
\begin{rema}\rm Notice that, if $\om$ is the volume form built by (\ref{vf}) from a Gindikin structure $\be^\la$ in the proof of the first part of the theorem, then the form $X_1^\la\wedge X_2^\la\wedge X_3^\la\iprod \om$, where $X_j^\la$ are given by (\ref{X`i}), coincides with $\be^\la$.

\end{rema}

\section{Appendix: Twisted metrics from $pp$-waves}
\label{apB}

Let $\theta(x,y,z,w)=F(y,w)$ be an obvious solution of II Plebański heavenly equation (\ref{IIPL}) (the so-called $pp$-wave). Then the corresponding SDVE metric
\begin{equation}\label{redmet}\equ
g= dw dx+dydz-F_{yy}dw^2
\end{equation}
has the Weyl spinor $8F_{yyyy}D_{X^1}^4$.
This solution satisfies system (\ref{IIH4D}) (the full Hirota system for II Plebański equation) if and only if
\begin{equation}\label{redHir}\equ
 yF_{yy}+wF_{wy}+F_{ww}=0.
\end{equation}
Below we give the general solution to this equation.

Introduce the following complex variable
\begin{equation*}
    \xi = \sqrt{4y-w^2} + i w, \qquad y>w^2/4.  \label{xi}
\end{equation*}
Using $\xi$ equation (\ref{redHir}) transforms to
\begin{equation}\equ
(\xi+\bar{\xi})F_{\xi\bar{\xi}}  = F_{\xi}+F_{\bar{\xi}} ,  \label{redHircom}
\end{equation}
which can also be written as
\begin{equation*}
  \Re(\xi F_{\xi\bar{\xi}}- F_{\xi}) =0 \qquad \text{or} \qquad
  \Re(\xi F_{\xi\bar{\xi}}- F_{\bar{\xi}}) =0 .
\end{equation*}
The second derivatives of $F$ with respect to $y$ and $w$ can be written using $\xi$:
\begin{equation*}
F_{yy}=\frac{16}{(\xi+\bar{\xi})^2}\Big(F_{\xi\xi}+F_{\bar{\xi}\bar{\xi}}\Big), \quad
F_{yw}=\frac{8i}{(\xi+\bar{\xi})^2}\Big(\xi F_{\xi\xi}-\bar{\xi}F_{\bar{\xi}\bar{\xi}}\Big), \quad
F_{yw}=-\frac{4}{(\xi+\bar{\xi})^2}\Big(\xi^2 F_{\xi\xi}+\bar{\xi}^2F_{\bar{\xi}\bar{\xi}}\Big),
\end{equation*}
where we used (\ref{redHircom}). It is easily seen that function
\begin{equation}\equ
  F(\xi,\bar{\xi})=\frac{1}{2}\big(\xi+\bar{\xi}\big)\big(\f'+\bar{\f}' \big)-\f-\bar{\f},
 \label{F}
\end{equation}
where $\f=\f(\xi)$ is an arbitrary holomorphic function of one variable, solves (\ref{redHir}).

Calculating
\begin{equation*}
   F_{yy}= \frac{8}{\xi+\bar{\xi}}\big(\f'''+ \bar{\f}'''\big)
\end{equation*}
allows us to reduce metric (\ref{redmet}) to
\begin{equation*}
 g=\frac{1}{4}\big[(\bar{\xi}dz-2idx)d\xi+(\xi dz+2idx)d\bar{\xi}\big]+\frac{2}{\xi+\bar{\xi}}\big[f(\xi)+\bar{f}(\bar{\xi})\big](d\xi-d\bar{\xi})^2,
\end{equation*}
where
\begin{equation*}
f(\xi)=\f'''(\xi).
\end{equation*}

The forth derivative of $F$ reads
\begin{equation*}
 F_{yyyy}=\frac{128}{(\xi+\bar{\xi})^3}\Bigg(f''+\bar{f}''-\frac{6(f'+\bar{f}')}{\xi+\bar{\xi}}+\frac{12(f+\bar{f})}{(\xi+\bar{\xi})^2} \Bigg).
\end{equation*}

The inverse to the twisted metric (see Theorem \ref{thTw}.2) in  the coordinates $(x,z,\xi,\xib)$ is given by
\begin{equation*}
  [g_\phi^{-1}]=-\frac{2}{(f+\fb)\phi'+(\xi+\xib)\phi}\begin{pmatrix}
     f+\fb   &  0   &   i\big(\xi-\frac{\phi'}{\phi}\fb\big)   & -i\big(\xib-\frac{\phi'}{\phi}f\big)\\
    * & 0 &  2  &   2 \\
   *  &  *  &  0  & -2\frac{\phi'}{\phi}(\xi+\xib)\\
 *  &   *  &  *  &  0
  \end{pmatrix}.
\end{equation*}

If the function $\phi(z)$ is nonconstant it is convenient to introduce a new coordinate
$Z$ by $z=\psi(Z):=\phi^{-1}(Z)$
and define a new function
$\Psi(Z):= \frac{\phi'(z(Z))}{Z}=\frac1{Z\psi'(Z)}\not=0$.
Then $dz=\frac{dZ}{Z \Psi(Z)}$
and the metric $g_\Psi:=g_\phi$ itself can be written as
\begin{equation*}
 g_\Psi= 2\big( \theta^1 \theta^4 - \theta^2\theta^3 \big),
\end{equation*}
where
\begin{align*}
\t^1 &= -\frac{1}{4}\Big(\frac{\xi\xib}{Z}dZ+\xib d\xi+\xi d\xib\Big),\\
\t^2 &=  \frac{2(\xi+\xib)\Psi dx-i(\xi^2-\xib^2+(\xi f-\xib \fb)\Psi)dZ/Z-i(\xi+\xib)(d\xi-d\xib)}{2(\xi+\xib+(f+\fb)\Psi)},\\
\t^3 &= Zdx+\frac{i(\xi\fb-\xib f)dZ+iZ(f+\fb)(d\xi-d\xib)}{2(\xi+\xib)},\\
\t^4 &= \frac{\xi+\xib+(f+\fb)\Psi}{(\xi+\xib)\Psi}dZ.
\end{align*}

Consider the simplest nontrivial particular case with the function $\f(\xi):=\xi^3/4$, i.e. $f(\xi)=3/2$, which gives the solution $F(y,w)=(4y-w^2)^{3/2}$ discussed in Example \ref{ppw}. Below we give formulas for the corresponding metric with arbitrary  $\Psi(Z)$, its Weyl spinor $C$ and fundamental invariants in the coordinates $(x,Z,\ka,\mu)$, where $\xi=\ka+i\mu$:
\begin{align*}
g=\frac2{2\ka +3\Psi}\left(-2\ka Z\Psi dx^2-2 \ka\mu\,  dx\, dZ-Z(2\ka -3\Psi) dx\,d\mu-\ka(4\mu^2+(2\ka +3\Psi)^2)dZ^2/{8Z\Psi}\right.\\
\left.-(2\ka +3\Psi)^2dZ\, d\ka/4\Psi-\mu (2\ka -3\Psi)dZ\, d\mu/2\Psi+3Zd\mu^2\right);\\
{\color{black}
C=\frac{4\Psi\big(3\Psi(6\kappa^2\mu^2\Psi\Delta_1-3\mu^4\Psi^2+(4\Delta_2-3\Delta_1^2)\kappa^4)+8\kappa^5\Delta_2 \big)}{Z^3(2\kappa+3\Psi)^5}D^4_{X^0}+
4\frac{36\big(\Delta_1\kappa^2-\mu^2\Psi\big)\mu\Psi^2}{Z^2\kappa(2\kappa+3\Psi)^4}D^3_{X^0}D_{X^1}}\\
{\color{black}
+6\frac{12\big(\Delta_1\kappa^2-3\mu^2\Psi\big)\Psi}{Z\kappa^2(2\kappa+3\Psi)^3}D^2_{X^0}D^2_{X^1}
-4\frac{36\mu\Psi}{\ka^3(2\ka +3\Psi)^2}D_{X^0}D^3_{X^1}-\frac{36Z}{\ka^4(2\ka +3\Psi)}D^4_{X^1};
}
\end{align*}
\begin{align}\equ
{\color{black}
I=\frac{1152\,\Psi\big(3(\Delta_1^2-\Delta_2)\Psi-2\kappa\Delta_2\big)}{Z^2(2\ka +3\Psi)^6}, \qquad
J=\frac{41472\,\Delta_1\Psi^2\big((2\Delta_1^2-3\Delta_2)\Psi-2\kappa\Delta_2)}{Z^3(2\ka +3\Psi)^9}, \label{IJ}
}\nonumber\\
I^3-6J^2=\frac{382205952\,\Delta_2^2\Psi^3\big((9\Delta_1^2-12\Delta_2)\Psi-8\kappa\Delta_2) \big)}{Z^6(2\ka +3\Psi)^{16}};
\end{align}
{\color{black}
here
$$
     \Delta_1:=2Z\Psi'(Z)+\Psi(Z), \qquad \Delta_2:=Z^2\Psi\Psi''+Z^2(\Psi')^2+4Z\Psi\Psi'+\Psi^2.
$$
}
The particular case $\Psi=1/Z$ corresponding to $\phi(z)=z$ (cf. Example \ref{ppw}) gives
$$
I=\frac{3456}{(2Z\ka+3)^6}, \qquad J=-\frac{82944}{(2Z\ka+3)^9},\qquad I^3-6J^2=0,
$$
{\color{black}while $\Psi=2/\sqrt{Z}$ (corresponding to $\phi(z)=z^2$) leads to $I=J=I^3-6J^2=0$ and $\Psi=3Z^{-1/3}$ (corresponding to $\phi(z)=z^3$) gives
$$
I=-\frac{3456(4\kappa Z^{1/3}+9)}{Z^{4/3}(2\kappa Z^{1/3}+9)^6}, \quad
J=-\frac{373248(4\kappa Z^{1/3}+12)}{Z^2(2\kappa Z^{1/3}+9)^9},
 \quad I^3-6J^2=-\frac{41278242816(16\kappa Z^{1/3}+45)}{Z^4(2\kappa Z^{1/3}+9)^{16}}.
$$
}
Taking $\Psi=1$ (corresponding to $\phi=e^z$) we get
$$
I=-\frac{2304\ka}{Z^2(2\ka+3)^6}, \quad 
J=-\frac{\color{black}41472(2\kappa+1)}{Z^3(2\ka+3)^9},
 \quad I^3-6J^2=-\frac{382205952(8\ka+3)}{Z^6(2\ka+3)^{16}}
$$
and for $\Psi=Z$:
$$
I=\frac{3456(3Z-4\ka){\color{black}Z}}{(2\ka+3Z)^6}, \quad J=-\frac{1492992\,Z^2\ka}{(2\ka+3Z)^9}, \quad  I^3-6J^2=\frac{41278242816\, Z^3 (3 Z-16 \ka)}{(2\ka+3Z)^{16}}.
$$
The Weyl tensor is algebraically special, i.e. $I^3=6J^2$, if and only if $\Delta_2=0$. The general solution
\begin{equation}\equ
   \Psi(Z)=\pm\frac{\sqrt{c_1 Z+c_2}}{Z}, \qquad c_1,c_2=const,  \label{AlSpSln}
\end{equation}
of the last equation leads to
$$
   I=\frac{3456\,c_2^2}{(2Z\kappa+3\sqrt{c_1Z+c_2})^6}, \qquad
   J=-\frac{82944\,c_2^3}{(2Z\kappa+3\sqrt{c_1Z+c_2})^9}.
$$
Thus (\ref{AlSpSln}) with $c_2\neq 0$ corresponds to algebraically special solutions  with $I\neq 0$ and $J\neq 0$. The deeper degenerations occur if and only if  $I=0$ if and only if $J=0$  if and only if $c_2=0$.


\providecommand{\bysame}{\leavevmode\hbox to3em{\hrulefill}\thinspace}
\providecommand{\MR}{\relax\ifhmode\unskip\space\fi MR }
\providecommand{\MRhref}[2]{%
  \href{http://www.ams.org/mathscinet-getitem?mr=#1}{#2}
}
\providecommand{\href}[2]{#2}

\end{document}